%% file: main.tex
\documentclass[10pt,journal,compsoc]{IEEEtran}
\usepackage{amsthm}
\usepackage{cite}
\usepackage{graphicx}
\usepackage{balance}  
\usepackage{color}
\usepackage{algorithmicx}
\usepackage{algpseudocode}
\usepackage{amsmath,amssymb,amsfonts,bm}
\usepackage{makecell}
\usepackage{subfigure}
\usepackage{multirow}
\usepackage[ruled,vlined]{algorithm2e}
\usepackage{booktabs}
\usepackage{hyperref}
\usepackage{threeparttable}
\usepackage{flushend}
\usepackage{upgreek}
\usepackage{array}
\usepackage{colortbl}
\usepackage{url}
\usepackage{enumitem}

\begin{document}
\title{MetaKG: Meta-learning on Knowledge Graph for Cold-start Recommendation}

\author{Yuntao~Du,
        Xinjun~Zhu,
        Lu~Chen,
        Ziquan~Fang,
        Yunjun~Gao,~\IEEEmembership{Member,~IEEE}

\IEEEcompsocitemizethanks{
        \IEEEcompsocthanksitem Y. Du, X. Zhu, L. Chen, Z. Fang, and Y. Gao (Corresponding Author) are with the College of Computer Science, Zhejiang University, Hangzhou 310027, China, E-mail:\{ytdu, xjzhu, luchen, zqfang, gaoyj\}@zju.edu.cn.
}

}

\IEEEtitleabstractindextext{
\begin{abstract}

A knowledge graph (KG) consists of a set of interconnected typed entities and their attributes. Recently, KGs are popularly used as the auxiliary information to enable more accurate, explainable, and diverse user preference recommendations. Specifically, existing KG-based recommendation methods target modeling high-order relations/dependencies from long connectivity user-item interactions hidden in KG. However, most of them ignore the cold-start problems (i.e., user cold-start and item cold-start) of recommendation analytics, which restricts their performance in scenarios when involving new users or new items. Inspired by the success of meta-learning on scarce training samples, we propose a novel meta-learning based framework called MetaKG, which encompasses a collaborative-aware meta learner and a knowledge-aware meta learner, to capture \textit{meta} users' preference and entities' knowledge for cold-start recommendations. The collaborative-aware meta learner aims to locally aggregate user preferences for each preference learning task. In contrast, the knowledge-aware meta learner is to globally generalize knowledge representation across different user preference learning tasks. Guided by two meta learners, MetaKG can effectively capture the high-order collaborative relations and semantic representations, which could be easily adapted to cold-start scenarios. Besides, we devise a novel adaptive task scheduler which can adaptively select the informative tasks for meta learning in order to prevent the model from being corrupted by noisy tasks. Extensive experiments on various cold-start scenarios using three real datasets demonstrate that our presented MetaKG outperforms all the existing state-of-the-art competitors in terms of effectiveness, efficiency, and scalability.

\end{abstract}

}

\maketitle

\IEEEdisplaynontitleabstractindextext

\IEEEpeerreviewmaketitle

\input{intro}

\input{relatedwork}

\input{background}

\input{formulation}

\input{methodology}

\input{complexity}

\input{experiment}

\input{conclusion}
\input{acknowledgments}

\balance

\bibliographystyle{abbrv}
\bibliography{refer}

\begin{IEEEbiography}[{\includegraphics[width=0.96in,height=1.1in,clip,keepaspectratio]
{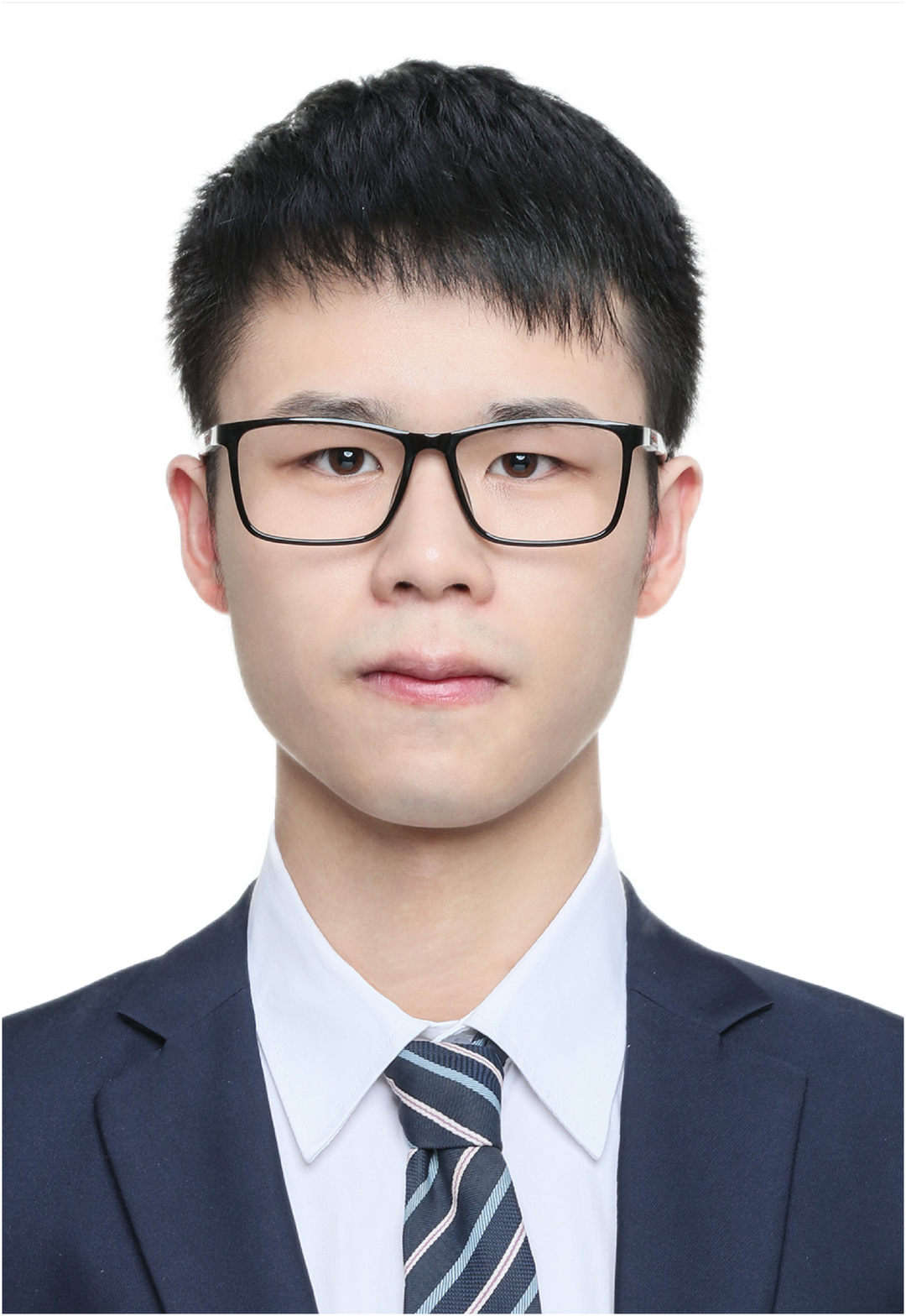}}]{Yuntao Du}
received the BS degree in data science from East China Normal University, China, in 2020. He is currently working toward the Master degree in the College of Computer Science, Zhejiang University, China. His research interests include data mining and recommender system.
\end{IEEEbiography}

\vspace*{-5ex}
\begin{IEEEbiography}[{\includegraphics[width=0.96in,height=1.1in,clip,keepaspectratio]
{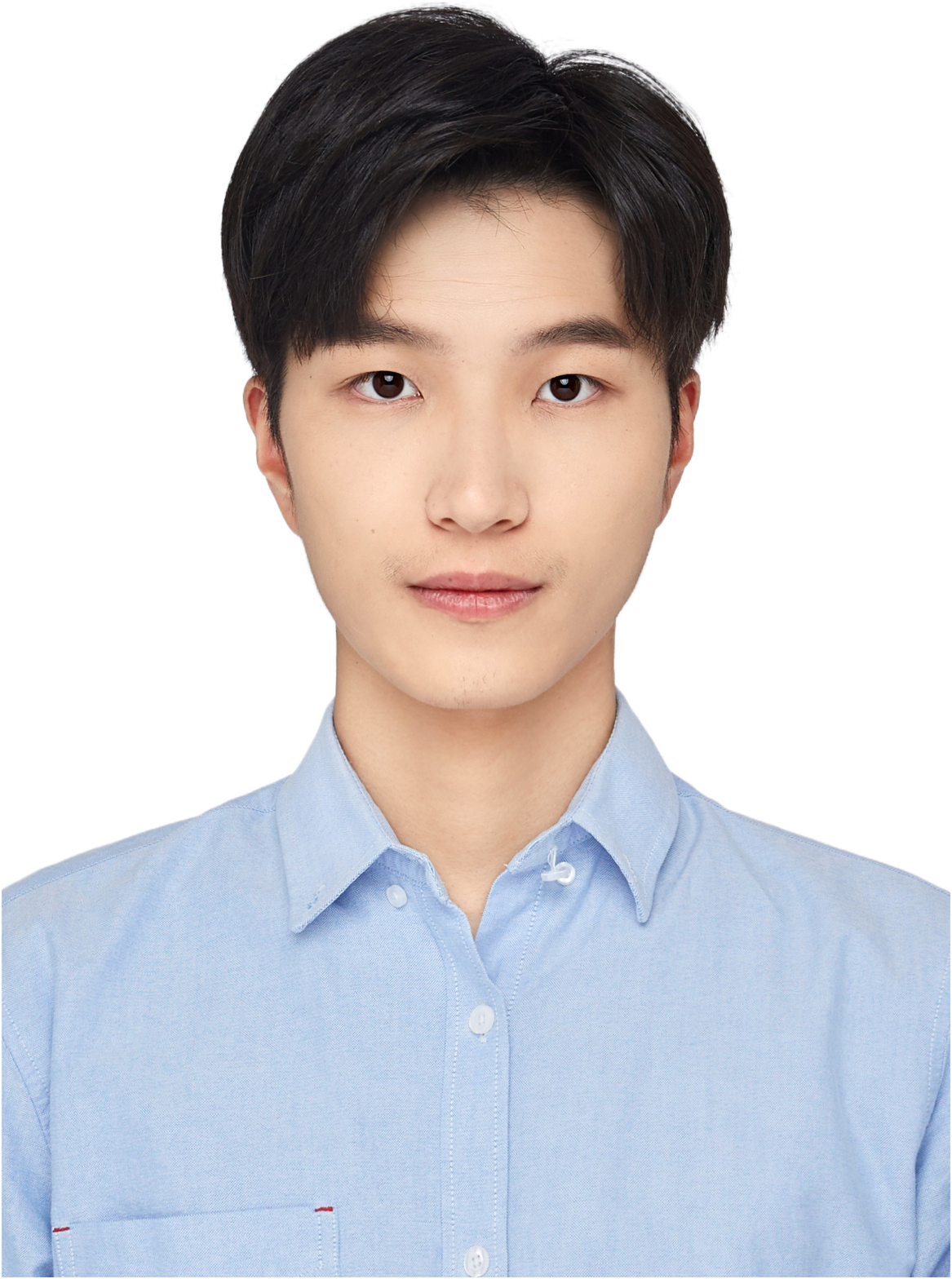}}]{Xinjun Zhu}
received the BS degree in computer science from Zhejiang University of Technology, China, in 2021. He is currently working toward the Master degree in the School of Software, Zhejiang University, China. His research interests include spatio-temporal data mining and recommender system.
\end{IEEEbiography}

\vspace*{-5ex}
\begin{IEEEbiography}[{\includegraphics[width=0.96in,height=1.1in,clip,keepaspectratio]
{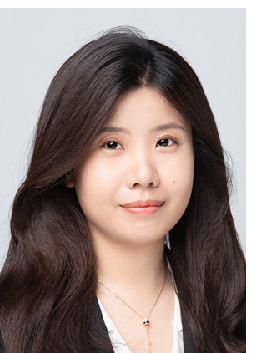}}]{Lu Chen}
received the PhD degree in computer science from Zhejiang University, China, in 2016. She was an assistant professor in Aalborg University for a 2-year period from 2017 to 2019, and she was an associated professor in Aalborg University for a 1-year period from 2019 to 2020. She is currently a ZJU Plan 100 Professor in the College of Computer Science, Zhejiang University, Hangzhou, China. Her research interests include indexing and querying metric spaces, graph databases, and database usability.
\end{IEEEbiography}

\vspace*{-5ex}
\begin{IEEEbiography}[{\includegraphics[width=0.96in,height=1.1in,clip,keepaspectratio]
{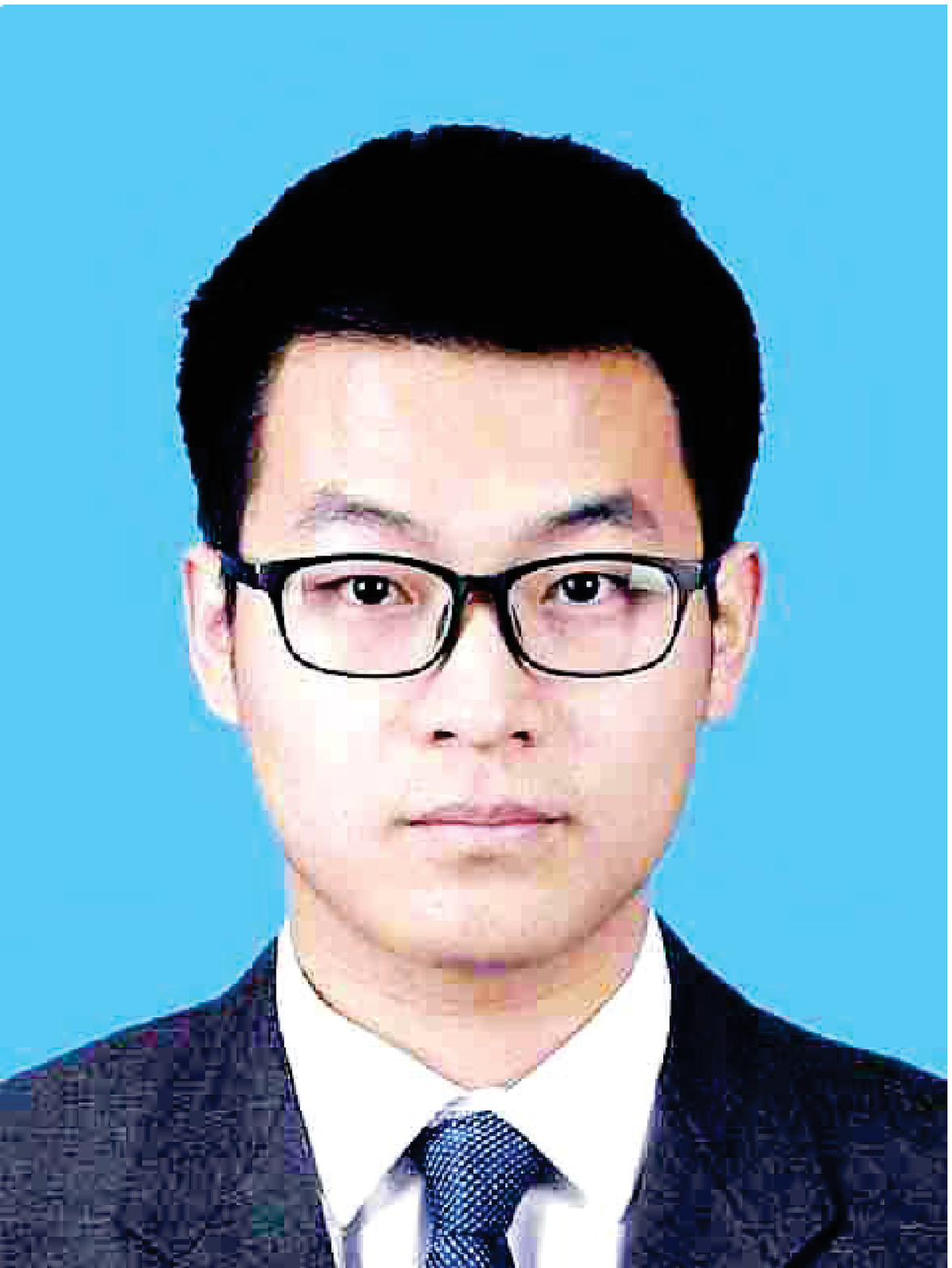}}]{Ziquan Fang}
received the BS degree in computer science from China Agricultural University, China, in 2018. He is currently working toward the PhD degree in the College of Computer Science, Zhejiang University, China. His research interests include trajectory data management and analytics.
\end{IEEEbiography}

\vspace*{-5ex}
\begin{IEEEbiography}[{\includegraphics[width=0.96in,height=1.1in,clip,keepaspectratio]
{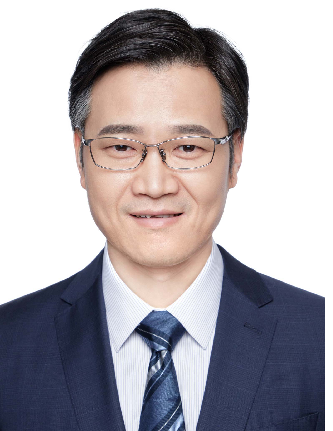}}]{Yunjun Gao}
received the PhD degree in computer science from Zhejiang University, China, in 2008. He is currently a professor in the College of Computer Science, Zhejiang University, China. His research interests include spatial and spatio-temporal databases, metric and incomplete/uncertain data management, graph databases, spatio-textual data processing, and database usability. He is a member of the ACM and the IEEE.
\end{IEEEbiography}

\end{document}

%% file: intro.tex
\IEEEraisesectionheading{\section{Introduction}\label{sec:introduction}}

\IEEEPARstart{W}{ith} the explosive use of the Internet, recommender systems are widely deployed in real-life applications like Twitter and Weibo, in order to provide personalized recommendation services while addressing information overload issues~\cite{recsys19basepaper, sigir19ngcf, kdd19kgat,wu2020garg}. \textit{User preference recommendation} aims to predict the preferences for target users based on the observed user-item interactions~\cite{icdm08cf, kdd08cf,YuQLL20}, which is a fundamental functionality among the current recommendation studies~\cite{dlrs16wide-deep}. Despite the efforts of traditional recommendation methods (e.g., collaborative-filtering-based methods~\cite{www17ncf} and content-based methods~\cite{www10content-rec}), they cannot mine the hidden high-order relations between users and items, thus restricting their effectiveness of user preference prediction. To this end, researchers start devoting knowledge graph (KG) based recommendations~\cite{www18dkn,cikm20mkgat}, which combines the user-item interactions with the side information of KG to improve the recommendation performance.

\begin{figure}[t]
\centering
\includegraphics[width=3.45in]{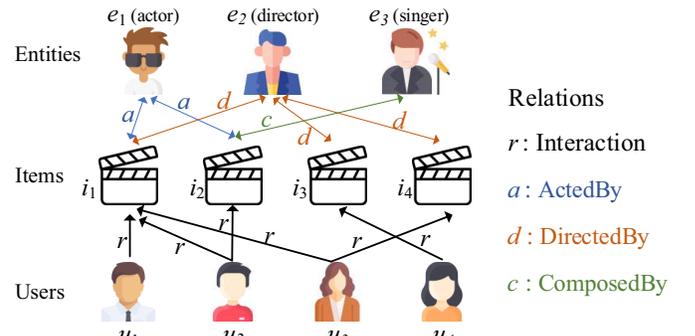}
\vspace{-5mm}
\caption{Knowledge Graph Based Movie Recommendation}
\label{fig:example}
\vspace{-4mm}
\end{figure}

Fig.~\ref{fig:example} gives an example of movie recommendation on KG that consists of users, items, and entities, where items denote movies to be recommended and entities (e.g, actors, directors, and singers) are the side information provided by KG. In Fig.~\ref{fig:example}, there exists an interaction relation between the user $u_1$ and the movie $i_1$, where $i_1$ is \textit{acted by} actor $e_1$ and is \textit{directed by} director $e_2$. Traditional recommendations only focus on the short-connectivity user-item relations such as $u \stackrel{r}{\rightarrow} i$. That is, they only capture the low-order dependences for recommendations, e.g., only users who have watched the same movies (e.g., $u_1$ and $u_2$ watch $i_1$) tend to share similar preferences. Hence,  the recommendation performance of traditional methods is limited. In contrast, the KG-based approaches are capable of discovering the long connectivity relations based on the side information hidden in KG, e.g., $u_1 \stackrel{r}{\rightarrow} i_1 \stackrel{d}{\rightarrow} e_2 \stackrel{d}{\rightarrow} i_3 \stackrel{r}{\rightarrow} u_4$. In that case, users $u_1$ and $u_4$ are also likely to share similar preferences even when they have not watched any same movies. 
Overall, the KG-based recommenders can fully exploit user-item interactions for better recommendation analytics.

Despite their effectiveness, we argue that current KG-based methods only focus on improving the recommendation performance from the data level (i.e., leveraging knowledge associations to provide additional semantics beyond collaborative signals), which would inevitably experience a lack of information and poor performance in cold-start scenarios because of overfitting and popularity biases~\cite{kdd19kgnn-ls, wsdm21coldstart}. Take the movie recommendation in Fig.~\ref{fig:example} as an example. If we want to recommend movies for user $u_4$, the recommendation results may be unsatisfied, as $u_4$ is only interacted with movie $i_3$. Based on the long connectivity path $u_4 \stackrel{r}{\rightarrow} i_3 \stackrel{d}{\rightarrow} e_2 \stackrel{d}{\rightarrow} i_1$ and the popularity of movie $i_1$ among existing users, the recommenders tend to recommend $i_1$ as the interested movie for $u_4$. However, another possible recommendation path $u_4 \stackrel{r}{\rightarrow} i_3 \stackrel{d}{\rightarrow} e_2 \stackrel{d}{\rightarrow} i_4$ is ignored due to insufficient interactions of movie $i_4$.

Recently, the success of meta-learning sheds light on modeling tasks with scarce training samples~\cite{maml, relationNN}. Specifically, meta-learning aims to learn prior knowledge from different tasks, which can be further employed to accommodate new tasks even when there are few data samples. Inspired by meta-learning, if we treat the preference learning for each user as a single task, we can model the preference learning for different users as different tasks under the meta-learning settings. Thus, the designed meta learner can derive \textit{meta} user-item interactions with strong generalization, which can be employed to deal with new tasks such as cold-start users or cold-start items. Although a few studies~\cite{kdd19metapred, kdd19metaLSTMRec} utilized meta-learning for cold-start problems and have achieved promising performance, they are traditional methods without any KG information leverage. Specifically, they either consider the short connectivity relations among users and items~\cite{kdd19melu} as discussed above, or rely on labor-intensive process to derive semantic facets~\cite{kdd20metaHIN}. Consequently, they cannot generalize generic prior knowledge for fast and accurate recommendations. To tackle the cold-start problems on KG-based recommendation, for the first time, we propose a meta-learning approach considering both user-item and KG. Two challenges exist as follows. 



\textbf{Challenge \uppercase\expandafter{\romannumeral 1}:} \textit{How to efficiently learn the user preference when evolving the cold-start issues?}
Existing KG-based recommendations highly rely on the massive user-item feedback, and thus, they
suffer from dramatic performance degradation when the user-item interactions are scarce~\cite{kdd19kgnn-ls, wsdm21coldstart}. This implies existing methods can not be directly adopted for our studied cold-start recommendations. Hence, it is necessary to redesign a learnable structure via meta-learning, which is able to capture the high-order \textit{meta} collaborative information from limited user-item interactions.


\textbf{Challenge \uppercase\expandafter{\romannumeral 2}:} \textit{How to capture entities' prior knowledge of KG to provide more accurate recommendation in cold-start scenarios?} Existing KG-based methods demonstrate that the semantic embeddings of entities from KG are beneficial for recommendation analytics~\cite{aaai19explainable, www19kgcn}. However, none of those methods consider how to derive \textit{meta} entities' knowledge with limited interactions. Thus, it is also challenging to make full use of the knowledge information (i.e., automatically learn strong generalizations for new users or new items) provided by KG under the meta-learning settings.

To address the above two challenges, we propose a meta-learning framework on KG for cold-start recommendation, called MetaKG, which simultaneously learns the general knowledge from collaborative relations and entities. Specifically, we first define the \textit{collaborative knowledge graph} (CKG)~\cite{kdd19kgat}, a hybrid structure of user-item interaction graph and knowledge graph, to represent the user behaviors and item knowledge in a unified way. Next, we design two novel meta learners to encode users' potential interests and entities' semantics from CKG, respectively.
The \textit{collaborative-aware} meta learner is developed to encode collaborative signals by exploiting gradients of aggregation layer within each user preference learning task, which can automatically generalize high-order connectivities between users and items.
The \textit{knowledge-aware} meta learner is further designed to capture the general knowledge semantics of KG by leveraging gradients of embedding and attention layers across tasks. 
In other words, the \textit{collaborative-aware} meta learner learns the user preference of each learning task to handle the first challenge, while the \textit{knowledge-aware} meta learner shares general semantic knowledge over different learning tasks to tackle the second challenge. 
Guided with two meta learners, MetaKG can achieve good generalization across tasks, and enables fast adaptation to cold-start scenarios with superior effectiveness and efficiency. To sum up, this paper makes the following key contributions:

\begin{itemize}[leftmargin=*]
    \item \textit{Meta-learning on KG.} We exploits meta-learning on KG-based recommendations, i.e., MetaKG, which achieves good generalization to alleviate the cold-start problems. To the best of our knowledge, this is the first attempt to investigate meta-learning on KG-based recommendations.
    \item \textit{Novel meta learners.} We design two meta learners, including \textit{collaborative-aware} meta learner and \textit{knowledge-aware} meta learner, to capture user preferences and knowledge associations. Thus, MetaKG can effectively capture the high-order collaborative relations and semantic representations with limited user-item interactions in cold-start scenarios.
    \item \textit{Adaptive task scheduler.} To provide informative tasks for meta learners and to prevent the model from being corrupted by noisy tasks, we further propose an adaptive task scheduler for the meta learner training process that decides which task to use later by predicting the probability being sampled for each task, and the scheduler is jointly optimized with two meta learners for better generalization.
    \item \textit{Extensive experiments.} We conduct extensive experiments on three public benchmarks while considering different cold-start scenarios. The results confirm that our proposed MetaKG substantially outperforms all the state-of-the-art methods.
\end{itemize}

The rest of this paper is organized as follows. Section~\ref{sec:relatedwork} reviews the related work, and Section~\ref{sec:background} presents the background. Section~\ref{sec:formulation} gives problem definitions. Section~\ref{sec:methodology} details our framework and methods. Section~\ref{sec:complexity} conducts theoretical analysis on computational complexity. The experimental results are reported in Section~\ref{sec:experiment}. Finally, Section~\ref{sec:conclusions} concludes the paper, and offers some research directions.

%% file: relatedwork.tex
\section{Related Work}\label{sec:relatedwork}
In this section, we review related work on KG-based recommendation and meta-learning, respectively.

\subsection{Knowledge Graph Based Recommendation}

Unlike the traditional methods that only consider the short connectivity user-item interactions for \textit{user preference recommendation}, more and more researchers start exploring knowledge graph (KG) based recommendation by combining user-item interactions with the side information 
to improve the recommendation performance. Here, we only highlight KG-based recommendation studies, while the comprehensive literature about traditional methods can refer to~\cite{RecSurvey19}.

Existing KG-based recommendation approaches can be mainly divided into three categories, i.e., embedding-based, path-based, and propagation-based methods. \emph{The embedding-based methods} directly embed entities and relations in KG via knowledge graph embedding (KGE) method~\cite{tkde17kgesuvey,recsys18rkge}. For example, CKE~\cite{kdd16cke} utilizes TransR~\cite{aaai15transR} to learn item embeddings of knowledge graph.
UGRec~\cite{sigir21UGRec} devises two different embedding spaces to model direct relations from KG and undirect co-occurrence relations among items.
However, the embedding-based methods simply adapt the KGE method into recommenders while ignoring the user-item interactions for recommendations. Thus, they cannot capture the complex dependencies of user-item relations for user preference learning. \emph{The path-based methods} aim to find different semantic paths in KG, and then connect items and users to guide recommendation process. For instance, MCRec\cite{kdd18metapath} develops deep neural networks with the co-attention mechanism to adopt rich meta-path-based context for top-$k$ recommendation. Similarly, KPRN~\cite{aaai19explainable} captures the sequential dependence within a path to generate path representations.
Nevertheless, the path-based approaches rely on handcraft meta-path construction, which is time-consuming and inevitably leads to information loss. \emph{The propagation-based methods} have attracted increasing attention in recent years, where the propagation process is performed iteratively to extract auxiliary information from KG for user preference learning. As an example,
KGAT~\cite{kdd19kgat} creates collaborative knowledge graph (CKG) to consider the user-item interactions and the side information of KG to enrich representations of users and items. CKAN~\cite{sigir20ckan} employs the heterogeneous propagation strategy to encode the user preference information with knowledge associations to further improve the recommendation performance. More recently, MKGAT~\cite{cikm20mkgat} integrates KG with additional image and text data to provide multi-modal knowledge for recommendation in an end-to-end fashion. Although existing KG-based methods achieve better performance than traditional methods, they still incur the cold-start problem that widely exists in real-world recommendation systems. For new users who have not yet interacted with enough items or new items that have not been visited before, existing KG-based methods fail to provide high quality recommendation results, since they heavily rely on abundant data to get desired performance, and inevitably lead to overfitting in cold-start scenarios. In light of this, for the first time, we apply a novel meta-learning approach to cold-start recommendation on KG, which can effectively derive prior collaborative signals and knowledge associations within and across different user preference learning tasks to support generic and accurate recommendations.

\subsection{Meta-learning}\label{sec:related_meta_learning}
Inspired by the human-learning, meta-learning~\cite{metalearning01} could learn the prior knowledge across different tasks and then quickly adapt the learned prior knowledge to new tasks. Specifically, meta-learning is often employed to learn a \textit{learner}, which can adapt to new tasks even when there is only a few training samples. Therefore, meta-learning has achieved great success in few-shot learning settings~\cite{nips16matchnet, nips17proto, maml, iclr17opti}. The existing work on meta-learning can be broadly divided into three categories: metric-based, memory-based, and optimization-based methods.

\emph{The metric-based methods} aim to learn a metric to evaluate the distance between instances. Siamese Network~\cite{SiameseNN} capitalizes on powerful discriminative features with a convolutional architecture and distance layer.
Prototypical Network~\cite{nips17proto} learns a metric space in which classification tasks can be performed by Euclidean distance computation to prototype representations of each class. However, the metric-based methods are mostly designed for classification problems, which are unsuitable for recommendation analytics involving prediction tasks. \emph{The memory-based methods} exploit memory architectures or specially-designed training processes to store key knowledge that can be generalized to new tasks. For example, MANN~\cite{MANN} introduces a memory-augmented model that can ingratiate significant short-term and long-term memory demands.
SNAIL~\cite{SNAIL} proposes a generic meta learner which combines temporal convolutions and soft attention to aggregate and memorize specific information for new unseen data. Nonetheless, these memory-based methods rely on extra memory mechanisms, resulting in too many parameters to tune. Finally, \emph{the optimization-based methods} learn parameters that are conducive to fast gradient-based adaptation to new tasks, which achieve the state-of-the-art performance. In this category,
MAML~\cite{maml} learns a good initialization from a base model. In that case, the adapted model can achieve strong generalization performance after applying a few gradient steps.
In this paper, we utilize the optimization-based meta-learning, as to be discussed in Section~\ref{sec:opt-meta}.

Inspired by the success of meta-learning, researchers start devoting efforts to address the cold-start recommendations~\cite{nips17metaRec, kdd19metaLSTMRec, kdd19melu}. MeLU~\cite{kdd19melu} uses the MAML framework~\cite{maml} for cold-start recommendations, and provides a candidate selection strategy for customized preference estimation. MetaHIN~\cite{kdd20metaHIN} proposes semantic-wise and task-wise adaptation to explore the semantics in heterogeneous networks. Moreover, specific scenarios (such as CTR prediction~\cite{sigir19ctr,www20metaselector} and clinical risk prediction~\cite{kdd19metapred}) are also investigated. However, these methods only consider \textit{one} meta leaner to learn \textit{one} optimization target (i.e., the rating loss), which are not feasible for KG-based recommendation since it is necessary to simultaneously capture the prior collaborative and knowledge information for user preference learning, and the simple model architectures of these methods are unable to capture the high-order relations between users and items.

%% file: background.tex
\section{Background}
\label{sec:background}

Graph neural networks (GNNs) have shown great success in various graph modelings, including directed/undirected graphs, heterogeneous graphs, and knowledge graphs. We design our base model by employing graph  attention networks to embed knowledge graphs, as an example of relation knowledge graph is shown in Fig.~\ref{fig:example}. Then, we utilize the optimization-based meta-learning to support the cold-start recommendations on embedded KG. In this section, we give a brief introduction to the GNNs and meta-learning related to the implementation of our presented framework. For ease of understanding, the frequently used notations are summarized in Table~\ref{tab:symbol}.

\begin{table}[t]
\centering
\caption{Symbols and Description}
\label{tab:symbol}
\setlength{\tabcolsep}{3pt}
\vspace{-3mm}
\begin{tabular}{|p{1.9cm}|p{6.5cm}|}
\hline
\textbf{Notation} & \textbf{Description} \\
\hline
$\mathcal{N}_{h}$ &  The set of triplets where $h$ is the head entity \\ \hline
$\mathcal{G}$ & The collaborative knowledge graph \\ \hline
$\mathcal{G}_k$, $\mathcal{G}_u$ & The knowledge graph, the user-item bipartite graph \\ \hline
$\mathcal{S}_u$ &  The support set of user $u$ \\ \hline
$\mathcal{Q}_u$ &  The query set of user $u$ \\ \hline
$\mathcal{T}_u = (\mathcal{S}_u, \mathcal{Q}_u)$ &  A rating task of user $u$ \\ \hline
$\delta$        &  The parameters of adaptive task scheduler \\ \hline
$\phi$          &  The parameters of embedding layer \\ \hline
$\omega$        &  The parameters of graph gated attention layer \\ \hline
$\gamma$        &  The parameters of aggregation layers \\ \hline
$v$ &  Learning rate of local update \\ \hline
$k$ &  Learning rate of global update \\ \hline
UC, IC , UIC &  user, item, user-item cold-start scenarios \\\hline
\end{tabular}
\vspace{-2mm}
\end{table}

\subsection{Graph Neural Networks (GNNs)}\label{sec:gnn}
Given a graph consisting of nodes and edges, GNNs aim to learn the representation for each node by leveraging the information propagation from its neighborhoods in both spectral and spatial domains. Meanwhile, GNNs perform aggregation operations from those learned embeddings. Specifically, GNNs capture the deep representation for each node $n_i$ by the following equations:
\begin{gather}
\mathbf{e}_{\mathcal{N}_i}^{l+1}=\text {PROP}^{l}\left(\left\{\!\!\left\{\left(\mathbf{e}_{i}^{l}, \mathbf{r}_{i, j},\mathbf{e}_{j}^{l}\right): j \in \mathcal{N}_i\right\}\!\!\right\}\right)\\
\mathbf{e}_{i}^{l+1}=\text{AGG}^{l}\left(\mathbf{e}_{i}^{l}, \mathbf{e}_{\mathcal{N}_i}^{l+1}\right)
\end{gather}
\noindent
Here, $\mathbf{e}_{i}^{l}$ represents the embedding of the $l$-th layer of node $n_i$, $\mathbf{r}_{i,j}$ stands for the representation of edge connecting node $n_i$ and node $n_j$, $\{\!\!\left\{\cdot\right\}\!\!\}$ denotes a multiset, and $\mathcal{N}_i$ is a set that represents the neighboring nodes of $e_i$. In addition, PROP denotes the information propagation operator, and AGG represents the information aggregation operator. When PROP propagates information from the neighboring edges and nodes to the source node, AGG transforms the entity embeddings into better ones at the same time. More details about GNNs can refer to~\cite{iclr18gat}.

There are lots of propagation methods to choose, including convolution~\cite{iclr17gcn}, self-attention~\cite{iclr18gat}, and knowledge-aware attention~\cite{kdd19kgat}. As for the aggregation methods, we can use GCN aggregator~\cite{iclr17gcn}, GraphSage aggregator~\cite{nips17graphSage}, and bi-interaction aggregator~\cite{kdd19kgat}. In this paper, we utilize knowledge-aware attention and bi-interaction aggregator for its simplicity and good performance.

\subsection{Optimization-based Meta-learning}\label{sec:opt-meta}
We employ the optimization-based meta-learning methods, which have been proved to achieve the state-of-the-art performance as discussed in Section~\ref{sec:related_meta_learning}. The optimization-based meta-learning aims at finding desirable parameters $\theta$ that are sensitive to new tasks, so that fine-tuning to those parameters could obtain large improvement in the task loss. Formally, given a distribution  $p (\mathcal{T})$ that all the tasks follow, the optimization-based meta-learning aims to train a model $f$ which is able to quickly adapt to new task $\mathcal{T}_{new}$ based on the knowledge obtained from $p (\mathcal{T})$. As shown in Fig.~\ref{fig:maml}, there are two kinds of update operations during the meta-learning process, namely, local update and global update. Specifically, the model locally updates its parameters $\theta$ to $\theta_{i}^{\star}$, which is performed though gradient descent updates on task $\mathcal{T}_{i}$ with a local update/learning rate $v$:
\begin{equation}
\theta_{i}^{\star}=\theta-v \nabla_{\theta} \mathcal{L}_{\mathcal{T}_{i}}\left(f_{\theta}\right)
\end{equation}
Based on the local update for task-specific parameter $\theta_{i}^{\star}$, the model further updates its global parameters by optimizing $\mathcal{L}_{\mathcal{T}_i} (f_{\theta_{i}^{\star}})$ over different tasks, with a global learning rate $k$. Thus, the global parameters $\theta$ can fit into various tasks.
\begin{gather}
\min _{\theta} \sum_{\mathcal{T}_{i} \sim p(\mathcal{T})} \mathcal{L}_{\mathcal{T}_{i}}\left(f_{\theta_{i}^{\star}}\right)=\sum_{\mathcal{T}_{i} \sim p(\mathcal{T})} \mathcal{L}_{\mathcal{T}_{i}}\left(f_{\theta-v \nabla_{\theta} \mathcal{L}_{\mathcal{T}_{i}}\left(f_{\theta}\right)}\right)\\
\theta \leftarrow \theta-k \nabla_{\theta} \sum_{\mathcal{T}_{i} \sim p(\mathcal{T})} \mathcal{L}_{\mathcal{T}_{i}}\left(f_{\theta_{i}^{\star}}\right)
\end{gather}
Note that, the optimization-based meta-learning divides each task into two sets for the model training, including the support set and the query set. During the local update, the model's local parameters $\theta^\star_{i}$ are trained over each support set, while the model searches task-specific desired optimization directions. During the global update, the model's global parameters $\theta$ (i.e., the directions in Fig.~\ref{fig:maml}) are altered based on the loss of query set using $\theta^\star_{i}$.

%% file: formulation.tex
\section{Problem Formulation}
\label{sec:formulation}

In this section, we proceed to introduce key concepts related to our studied problem, and present problem formulation.

\begin{figure}[tb]
	\centering
	\includegraphics[width=0.45\textwidth]{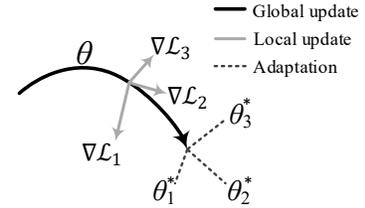}
	\vspace{-5mm}
	\caption{Diagram of Optimization-based Meta-learning}
	\label{fig:maml}
	\vspace{-3mm}
\end{figure}

\textbf{User-Item Bipartite Graph.} In recommendations, the core task is how to represent the user-item relations. We model the user-item interactions with a user-item bipartite graph $\mathcal{G}_u$, defined as follows:
\begin{gather}
\mathcal{G}_u  = \{(u,i,l)\ | u \in \mathcal{U},  i \in \mathcal{I} ,l_{ui} \in \mathcal{L}\}
\end{gather}
where $\mathcal{U}$ and $\mathcal{I}$ denote the user set and item set, respectively. $\mathcal{L} = \mathcal{U} \times \mathcal{I} $ defines the user-item edges/interactions. In $\mathcal{G}_u$, if there is an observed edge between user $u$ and item $i$, $l_{ui} = 1$; otherwise $l_{ui} = 0$.

\textbf{Knowledge Graph.} To capture the high-order relations between uses and items discussed in Fig.~\ref{fig:example}, we introduce knowledge graph (KG) to provide the side information. A KG is a directed graph $\mathcal{G}_k$ as defined below:
\begin{gather}
\mathcal{G}_k  = \{(h,r,t)\ | h\in \mathcal{E}, t \in \mathcal{E},  r \in \mathcal{R}\}
\end{gather}
Here, each triple $(h, r, t)$ in $\mathcal{G}_k$ means that there is a relation $r$ between the head entity $h$ and the tail entity $t$. $\mathcal{E}$ and $\mathcal{R}$ represent the entity set and the relation set in $\mathcal{G}_k$, respectively. For example, a triple (\textit{Mark Hamill}, \textit{ActorOf}, \textit{Star War}) indicates that \textit{Mark Hamill} is an actor of the movie \textit{Star War}. We assume the relations in $\mathcal{R}$ are bidirectional, so that they contain both canonical direction (i.e. \textit{ActorOf}) and inverse direction (i.e. \textit{ActedBy}).

Based on the defined user-item bipartite graph ${G}_u$ and knowledge graph ${G}_k$, we can further define a set $\mathcal{A} = \{(i, e)\ | i \in \mathcal{I}, e \in \mathcal{E}\}$, where $\mathcal{I}$ belongs to $\mathcal{G}_u$, $\mathcal{E}$ belongs to $\mathcal{G}_k$, and $(i,e)$ means that the item $i$ from $\mathcal{G}_u$ can be aligned with the entity $e$ (i.e., $h$ or $r$) from $\mathcal{G}_k$.

\begin{figure*}[tb]
	\centering
	\vspace{-5mm}
	\includegraphics[width=1.0\textwidth]{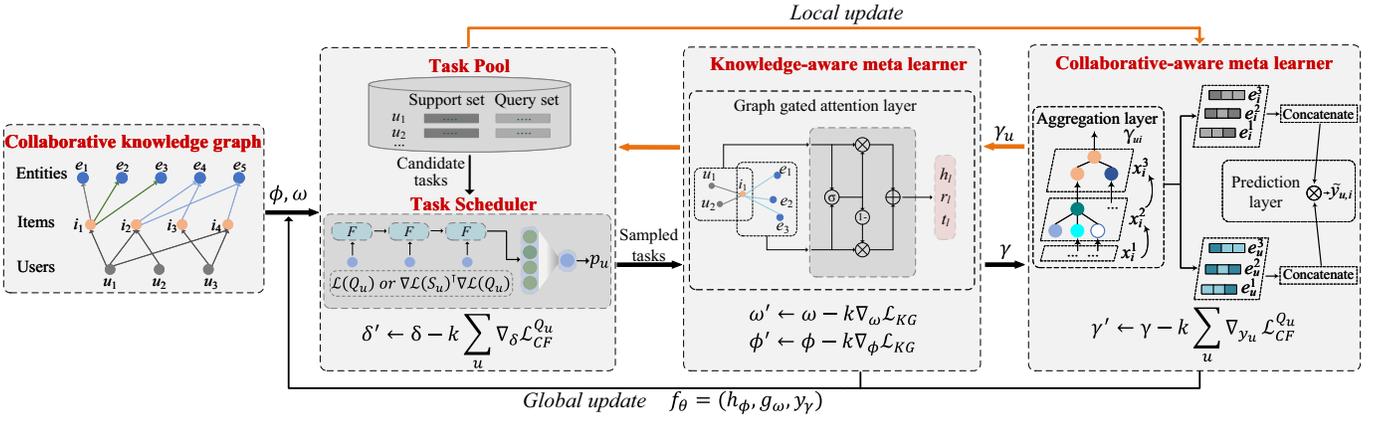}
	\vspace{-5mm}
	\caption{An Overview of MetaKG Framework}
	\label{fig:overview}
	\vspace{-3mm}
\end{figure*}

\textbf{Collaborative Knowledge Graph.} Similar to the previous studies on KG-based recommendations~\cite{kdd19kgat}, we define the collaborative knowledge graph (CKG) $\mathcal{G}$ as a relational graph that combines user-item bipartite graph $\mathcal{G}_u$ and knowledge graph $\mathcal{G}_k$ in a unified manner. Specifically, we integrate $\mathcal{G}_u$ with $\mathcal{G}_k$ based on the item-entity alignment set $\mathcal{A}$. Overall, the final CKG $\mathcal{G}$ is defined as:
\begin{gather}
\mathcal{G}  = \{(h,r,t)\ | h \in \mathcal{E^\prime} ,t \in \mathcal{E^\prime},  r \in \mathcal{R^\prime}\}
\end{gather}
Here, $\mathcal{E^\prime} = \mathcal{E} \cup \mathcal{U}$, indicating that we treat users as entities as well. Besides, $\mathcal{R^\prime} = \mathcal{R} \cup \mathcal{L}$, meaning that we treat user-item interactions as relations as well.

\textbf{Problem Statement.} Given a collaborative knowledge graph $\mathcal{G}$ that combines the user-item bipartite graph $\mathcal{G}_u$ and the knowledge graph $\mathcal{G}_k$, we aim to predict the unknown probability (i.e., user preference) $p_{ui}$ from user $u$ to item $i$, where $r_{u,i} \notin \mathcal{R^\prime}$. Specifically, if $u$ is a new user with only a small number of interactions, i.e., $|r_{u^\prime,i} \in \mathcal{R}^\prime: u^\prime = u|$ is small, it is known as \underline{u}ser \underline{c}old-start problem (UC). Similarly, if $i$ is a new item with limited interactions from existing users, i.e., $|r_{u,i^\prime} \in \mathcal{R}^\prime: i^\prime = i|$ is small, it is known as \underline{i}tem \underline{c}old-start problem (IC). Last but not the least, if both $u$ and $i$ are new, it is known as \underline{u}ser-\underline{i}tem \underline{c}old-start problem (UIC).

In this paper, our objective is to alleviate the cold-start problems involving UC, IC, and UIC perspectives using meta-learning on KG-based recommendations.

%% file: methodology.tex
\section{Methodology}\label{sec:methodology}

In this section, we first give an overview of the MetaKG framework, and then, we present the detailed methods.

\subsection{Overivew of MetaKG}
We propose the MetaKG framework in both architecture and learning perspectives to better illustrate each module's functionality as well as the meta-learning process.

In terms of the architecture perspective, MetaKG is established on three essential components: adaptive task scheduler, \textit{collaborative-aware} meta learner and \textit{knowledge-aware} meta learner, as depicted in Fig.~\ref{fig:overview}. The adaptive task scheduler (to be discussed in Section~\ref{sec:Adaptive task scheduler}) selects the most informative tasks from the task pool to filter out noisy tasks for meta learners training. The collaborative-aware meta learner (to be discussed in Section~\ref{sec:collaborative-meta learner}) aggregates hidden information with user preference in the aggregation layer in order to capture the high-order relations. It is worth mentioning that the collaborative-aware meta learner is able to automatically learn  high-order collaborative signals, without relying on the predefined meta-paths or domain knowledge~\cite{kdd20metaHIN}. In addition, we further design the knowledge-aware meta learner (to be discussed in Section~\ref{sec:knowledge-meta learner}) to capture the general knowledge semantics shared by tasks. Equipped with two meta learners, MetaKG enables to explicitly encode the collaborative signals and knowledge associations in a meta-learning manner, and achieves fast adaptation for new users and new items.


In terms of the learning perspective, MetaKG consists of three procedures: local update (the orange lines at the top in Fig.~\ref{fig:overview}), global update (the black lines at the bottom in Fig.~\ref{fig:overview}), and adaptation process. During the local update, the collaborative-aware meta learner is utilized to learn the personalized preference for each task/user. During the global update, the knowledge-aware meta learner is further employed to learn general meta knowledge associations of KG shared across different tasks/users. Once our model has been well-trained, it can easily adapt to the new scenarios with only a few gradient descent steps, and achieve high quality predictions, as indicated in our experiments.

\subsection{Meta-learning Setups}
Since MAML~\cite{maml} is originally designed as a model-agnostic meta-learning framework, it is necessary to clarify more setups when adapting it to the cold-start recommendations. We define a task for a user $u$ as $\mathcal{T}_u = (\mathcal{S}_u,\mathcal{Q}_u)$, indicating that one task is bound to one user. Here, $\mathcal{S}_u$ denotes the support set and $\mathcal{Q}_u$ represents the query set, as discussed in Section~\ref{sec:opt-meta}. During the training/learning process, $\mathcal{Q}_u$ and $\mathcal{S}_u$ are sampled from the item set
which a user $u$ has interacted with. Note that the support set and query set are mutually exclusive. During the adaptation process, each task $\mathcal{T}_u \in \mathcal{T}_{test}$ also has a support set and a query set, and items in the query set are to be predicted. As defined in Section~\ref{sec:formulation}, \textbf{UC} denotes the task that new users are not seen in training process, \textbf{IC} stands for the task which new items are not seen in training process, and \textbf{UIC} represents the task that both users and items are not seen in training process.

\subsection{Base Model}\label{sec:base model}
As shown in Fig.~\ref{fig:overview}, we design the base model $f_{\theta}$ that contains four layers: i) embedding layer $h_{\phi}$, which parameterizes the entities and relations as the vector representations; ii) graph gated attention layer $g_{\omega}$, which utilizes relation-aware gated attention mechanism to neighborhood nodes; iii) aggregation layer $y_{\gamma}$, which aggregates entity representation with its ego-network information; and iv) prediction layer, which predicts a matching score based on the user and item representations. Overall, the proposed base model can be represented $f_{\theta} = (h_{\phi}, g_{\omega}, y_{\gamma})$, as described below.

\textbf{(i) Embedding layer.} Both entity embeddings and relation embeddings are important to preserve the graph structure of knowledge graph. Hence, we first utilize TransR~\cite{aaai15transR} on CKG to get the initial embeddings. Then, we employ a learnable projection layer to enrich their representations:
\begin{gather}
    e_{i} = h_{\phi}(c_i) = W_{e} c_i + b
\end{gather}
where $c_i$ is a $d_p$-dimensional one-hot vector denoting entity or relation of CKG, and $W_{e} \in \mathbb{R}^{d_e\times d_p}$ and $b \in \mathbb{R}^{d_e}$ are trainable parameters, i.e., $\phi = \{W_{e}, b\}$.

\textbf{(ii) Graph gated attention layer.} Previous methods~\cite{kdd19kgat,sigir20ckan} typically aggregate all neighbor triplets to generate the contextual representations, without differentiating the semantics that neighbor entities carry with. For example, for item $h$, the neighbor entity $t$ and neighbor user $u$ represent the knowledge relevance and collaborative connectivities of $h$, respectively. However, directly aggregating them would fail to generate discriminative signals towards user’s preference representations, thus leading to suboptimal performance. Therefore, we aim to design a gated attention module for explicitly aggregating two kinds of information. Specifically, we use $\mathcal{N}_h^c = \{(h,r,u) | (h,r,u) \in \mathcal{G}, u\in \mathcal{U}\}$ to represent the set of triplets where $h$ is the head entity and the tail entity $u$ is an user, and summarize the weighted representation to capture the first-order information:
\begin{gather}
\overrightarrow{e_h^c}=g_{\omega} (h, r, u)= \sum_{(h, r, u) \in \mathcal{N}_{h}^c} \alpha(h, r, u) e_{u}
\end{gather}
where $\alpha(h, r, u)$ is the attention weight, indicating the importance of tail $u$ and relation $r$ to head $h$. We use relation-aware attention mechanism to calculate such importance:
\begin{gather}
\alpha^\prime(h, r, u)=(W_{r} e_{u})^{\top} \tanh ((W_{r} e_{h} + e_{r})\\
\alpha(h, r, u)=\text{softmax}(\alpha^\prime(h,r,u))
\end{gather}
here, $W_r \in \mathbb{R}^{d_r\times d_e}$ projects entities from the $d_e$-dimension entity space into the $d_r$-dimension relation space. We use the same attention strategy to generate the $\overrightarrow{e_h^k}$ for entity $h$ with knowledge neighbor triplets $\mathcal{N}_h^k = \{(h,r,t) | (h,r,t) \in \mathcal{G}, t\in \mathcal{E}\}$. Inspired by the design of GRU~\cite{14gru} that learns gating signals to control the update of hidden states, we propose to learn a fusion gate which can adaptively control the combination of two semantic representations:
\begin{gather}
    g_h = \sigma(W_c \overrightarrow{e_h^c} + W_k \overrightarrow{e_h^k}) \\
    \overrightarrow{e_h} = g_h\cdot \overrightarrow{e_h^c} + (1-g_h) \overrightarrow{e_h^k}
\end{gather}
where $W_c, W_k \in\mathbb{R}^{d_e\times d_e}$ are learnable transformation parameters and $\sigma$ is the sigmoid function. The $g_h\in\mathbb{R}^e$ denotes the learned gate signals to balance the contributions of collaborative signals and knowledge associations, and the gated attention layer is parameterized by $\omega = \{W_{r}, W_c, W_k\}$.

\vspace{1mm}
\textbf{(iii) Aggregation layer.} After finishing information propagation, a bi-interaction aggregator is used to model two kinds of feature interaction between $e_h$ and $\overrightarrow{e_h}$ as follows:
\begin{equation}
\begin{aligned}
e^{l+1}_h = y_{\gamma}^l(e_h^{l},\overrightarrow{e_h}^{l})=& \text { LeakyReLU }\left(W_{1}^{l}(e_{h}^{l}+\overrightarrow{e_h}^{l}))\right)+\\
& \text { LeakyReLU }\left (W_{2}^{l}(e_{h}^{l} \odot \overrightarrow{e_h}^{l} )\right)
\end{aligned}
\end{equation}
where $W_1, W_2 \in \mathbb{R}^{d_l \times d_e}$ are learnable matrices, $\odot$ denotes the element-wise product, and $l$ is the $l$-layer propagation. Therefore, $y_{\gamma}$ is the aggregative function parameterized by $\gamma = \{(W_{1}^{l}, W_{2}^{l})\ | \forall l\in\{1,2,...,L\} \}$.

\vspace{1mm}
\textbf{(iv) Prediction layer.} In terms of preference prediction, given user's multi-layer representation $\{e_{u}^1,...,e_{u}^L\}$ and item's multi-layer representation $\{e_{i}^1,...,e_{i}^L\}$, we concatenate each layer representation into a single vector:
\begin{gather}
e_{u}^{\star}=e_{u}^1\|\cdots\| e_{u}^L, \quad e_{i}^{\star}=e_{i}^1\|\cdots\| e_{i}^L
\end{gather}
Then, we predict their preference score by inner product of user and item representations, and opt for the BPR loss~\cite{uai09BPRloss} to learn the preference of users below:
\begin{equation}
\mathcal{L}=\sum_{(u, i, j) \in \mathcal{O}}-\ln \sigma({e^{\star}_{u}}^{\top}{e^{\star}_{i}}-{e^{\star}_{u}}^{\top}{e^{\star}_{j}})
\end{equation}
where $\mathcal{O} = \{(u,i,j)\ | (u,i) \in \mathcal{O}^{+},(u,j) \in \mathcal{O}^{-}\}$ is the training dataset consisting of the observed interactions $\mathcal{O}^{+}$ and unobserved counterparts $\mathcal{O}^{-}$; and $\sigma(\cdot)$ is the sigmoid function. It is worth mentioning that the base model is supervised, and needs a lot of data to get reasonable performance, which is not feasible for cold-start scenarios. Thus, we regard the cold-start recommendation as a meta-learning problem. Different from previous meta-learning approaches~\cite{kdd19melu, kdd20metaHIN} that consider \textit{one} meta learner to learn \textit{one} optimization target (\textit{i.e.}, the rating loss), MetaKG includes two novel meta learners with different optimization targets to encode the prior information from both user-item interactions and KG. Next, we detail our proposed adaptive task scheduler and meta learners as well as the co-adaptation process.

\subsection{Adaptive Task Scheduler}\label{sec:Adaptive task scheduler}
Previous meta-learning based recommendations~\cite{kdd19melu,kdd20metaHIN} randomly sample meta-training tasks with a uniform probability, under the assumption that tasks are of equal importance. However, given rather limited number of meta-training tasks in cold-start scenarios, it is likely that some tasks are noisy~\cite{icdm08cf} or contribute differently for user preference learning. To prevent the meta learners from being corrupted by such detrimental tasks or dominated by tasks in the majority, we propose an adaptive task scheduler to select the most informative tasks for training. Specifically, we define the scheduler as $z$ with parameters $\delta$, and utilize two representative factors to quantify the information covered in $\mathcal{T}_u$:
\begin{equation}\label{equ:scheduer}
p_u = z_\delta \big(\mathcal{L}(\mathcal{Q}_u), \nabla\mathcal{L}(\mathcal{S}_u)^{\top}\nabla\mathcal{L}(\mathcal{Q}_u)\big)
\end{equation}
where $p_u$ is the sampling probability of candidate task $\mathcal{T}_u$, $\mathcal{L}(\mathcal{Q}_u)$ denotes the loss on the query set, and the second part is the gradient similarity between the support and query set of task $\mathcal{T}_u$. These factors are associated with the learning outcome and the learning process of the task $\mathcal{T}_u$, and a large query set loss may represent at true hard task if the gradient similarity is large; a task with noises in its query set could lead to large query loss but small gradient similarity. Thus, we follow~\cite{nips21scheduler} to choose LSTM network~\cite{97lstm} as the scheduler $z$ to encode the historical information of each task and to capture the prediction variance~\cite{nips17high_variance}. The parameters $\delta$ of the scheduler are simultaneously optimized with the model, as demonstrated in Algorithm~\ref{alg:MetaKG}.

\subsection{Collaborative-aware Meta Learner}\label{sec:collaborative-meta learner}
The collaborative-aware meta learner aims to learn multifaceted preference from CKG. Here, we also use the BPR loss~\cite{uai09BPRloss} to learn the preference of each user:
\begin{equation}
\mathcal{L}_{\mathrm{CF}}^{\mathcal{S}_u}=\sum_{(u, i, j) \in \mathcal{S}_u}-\ln \sigma({e^{\star}_{u}}^{\top}{e^{\star}_{i}}-{e^{\star}_{u}}^{\top}{e^{\star}_{j}})
\end{equation}
where $\mathcal{S}_u = \{(u,i,j)\ | (u,i) \in \mathcal{O}^{+},(u,j) \in \mathcal{O}^{-}\}$ is the support set for task $u$. Given a task $\mathcal{T}_u$, the user's embedding $e_u$, and the weighted propagation representation $\overrightarrow{e_u}$ at $l$-layer, we encode feature interaction at each layer, and concatenate into one final representation as follows:
\begin{equation}
e^{\star}_u = {||}_{l\in L}y_{\gamma}^l(e_u^l,\overrightarrow{e_u}^l)
\end{equation}

In addition, the collaborative signals are automatically captured by collaborative-aware meta learner through learnable interaction aggregator $\gamma$ of task $\mathcal{T}_u$:
\begin{equation}
\gamma_u = \gamma - v \frac{\partial \mathcal{L}_{\mathrm{CF}}^{\mathcal{S}_u}(\phi,\omega, e_u^\star)}{\partial \gamma}
\end{equation}
where $v$ is the learning rate of local update, and $\gamma_u$ is specific information aggregator of task $\mathcal{T}_u$. Therefore, collaborative-aware meta learner locally updates the aggregation layer by optimizing $\mathcal{L}_{\mathrm{CF}}$ on the support set of each user task in order to capture the user-specific preference.

\subsection{Knowledge-aware Meta Learner}\label{sec:knowledge-meta learner}
The knowledge-aware meta leaner models the entity representation on the granularity of triplets. We first introduce the energy score for a given triplet $(h,r,t)$:
\begin{equation}
s(h, r, t)=\|W_{r} e_{h}+e_{r}-W_{r} e_{t}\|_{2}^{2}
\end{equation}
where $e_{h}, e_{r},$ and $e_{t}$ are the embeddings of the head entity, the relation, and the tail entity, respectively. This means we first project the entities into the relation space with the help of $W_r$ (i.e., $W_{r}e_{h}$ and $W_{r} e_{t}$), and then model the additive relationship of knowledge triplet~\cite{aaai15transR}. The smaller the energy score $s(h, r, t)$ is, the more likely the triplet $s(h, r, t)$ is true. Based on this, a pairwise ranking loss is used to encourage the direct connections through negative sampling:
\begin{equation}
\mathcal{L}_{\mathrm{KG}}=\sum_{\left(h, r, t, t^{\prime}\right) \in \mathcal{S}}-\ln \sigma\left(s\left(h, r, t^{\prime}\right)-s(h, r, t)\right)
\end{equation}
Here, $\mathcal{S} = \{((h, r, t, t^{\prime})\ | (h,r,t)\in \mathcal{G}, (h,r,t^{\prime}) \notin \mathcal{G}\} $, and $\sigma(\cdot)$ is the sigmoid function.

Recall that the entity representations are parameterized by $\phi$ and $\omega$. The semantic information of entities should not be adapted with a specific task, but should be shared across the whole knowledge graph. Motivated by this, we design a knowledge-aware meta learner to learn knowledge associations and semantic relatedness over tasks:
\begin{gather}
\phi^\prime = \phi - k \frac{\partial \mathcal{L}_{\mathrm{KG}}(\gamma,c)}{\partial \phi} \\
\omega^\prime = \omega - k \frac{\partial \mathcal{L}_{\mathrm{KG}}(\gamma,c)}{\partial \omega}
\end{gather}
where $k$ is the learning rate of global update and $c$ is the input one-hot vector of entities. In other words, knowledge-aware meta learner globally updates the embedding layer and attention layer by optimizing $\mathcal{L}_{\mathrm{KG}}$ across all the tasks.

\begin{algorithm}[t]
\LinesNumbered
\DontPrintSemicolon
\caption{Meta Learning on knowledge graph for cold-start recommendation}
\label{alg:MetaKG}
    \KwIn{$\mathcal{G},\mathcal{T},\theta,\omega,\phi,\gamma,\delta,c,k$}
    \KwOut{the learned parameters $\theta^\prime$ in a new scenario}
    Initialize entity embedding with TransR on $\mathcal{G}$\;
    \While {not converge}{
        Calculate the sampling probability $p$ of each task in $\mathcal{T}_{train}$ by Eq.~\ref{equ:scheduer} \;
        Sample batch of task $(\mathcal{S},\mathcal{Q})\in \mathcal{T}_{train}$ according to the sampling probabilities \;
        \For {$\mathcal{S}_u \in \mathcal{S}, \mathcal{Q}_u \in \mathcal{Q}$} {
            Calculate user representation $e^\star_u$\ through $\theta$\;
            Local update $\gamma_u \leftarrow \gamma - v \nabla_{\gamma} \mathcal{L}_{\mathrm{CF}}^{\mathcal{S}_u}$\;
        }
        $\begin{aligned} \text { Global update } & \omega^{\prime} \leftarrow \omega - k \nabla_{\omega}\mathcal{L}_{\mathrm{KG}} \\ & \phi^{\prime} \leftarrow \phi - k \nabla_{\phi}\mathcal{L}_{\mathrm{KG}} \\ &  \gamma^\prime \leftarrow \gamma - k \sum_{u} \nabla_{\gamma_u}\mathcal{L}_{\mathrm{CF}}^{\mathcal{Q}_u} \end{aligned}$ \;
        Update the scheduler $\delta^\prime \leftarrow \delta - k\sum_{u} \nabla_{\delta}\mathcal{L}_{\mathrm{CF}}^{\mathcal{Q}_u}$
    }
Co-adapt to the new scenario according to Eq.~\ref{equ:adaptation} \;
\Return{$\theta^\prime$}
\end{algorithm}

\subsection{Co-adaptation}
Our MetaKG learns the model parameters $\theta$ based on the user preference and the knowledge associations by optimizing two meta learners alternatively. During the training process, the collaborative-aware meta learner locally updates aggregation layer $\gamma$ through backpropagation of prediction loss according to Eq. 21, while the knowledge-aware meta learner globally optimizes the embedding layer $\phi$ and attention layer $\omega$ through backpropagation of KG loss according to Eq. 24 and 25. Note that, we first train our model to get the meta parameters $\theta^\prime$, and then, we adapt $\theta^\prime$  to new scenarios (e.g., user cold-start) as below:
\begin{equation}\label{equ:adaptation}
\min_{\theta^\prime}  \sum_{\mathcal{T}_{u} \in \mathcal{T}_{new}} \mathcal{L}^{\mathcal{S}_u}_\mathrm{CF}(\phi^\prime,\omega^\prime, e^{\star}_u) + \mathcal{L}_\mathrm{KG}(\gamma^\prime,c)
\end{equation}
where $\mathcal{T}_{new}$ denotes the new scenarios; while $\phi^\prime$, $\omega^\prime$, and $\gamma^\prime$ are learned parameters from two meta learners during local and global updates.
During the adaptation process, MetaKG further updates the learned parameters using the support set of a new scenario. The detailed training process of MetaKG is presented in Algorithm~\ref{alg:MetaKG}.

%% file: complexity.tex
\section{Model Analysis}\label{sec:complexity}
In this section, we present the theoretical analysis on the computational complexity and the model parameters.

\textbf{Computational complexity.} The time complexity of MetaKG for recommendation involves three aspects. For the adaptive task scheduler, the task probability estimation has the computational complexity of $\mathcal{O}(|\mathcal{T}|d_{lstm}^2)$, where $|\mathcal{T}|$ is the number of tasks, and $d_{lstm}$ is the hidden size of LSTM. For the collaborative-aware meta-learner, the matrix multiplication of the $l$-th layer has the computational complexity of $\mathcal{O}(|\mathcal{G}| d_{l} d_{l-1}) $, where $|\mathcal{G}|$ is the number of triplets in CKG, and $d_l$ and $d_{l-1}$ denote the current and previous transformation size. For the knowledge-aware meta learner, the knowledge graph embedding has the computational complexity of $\mathcal{O}(|\mathcal{G}_k| d_{p} d_{e})$, where $d_{p}$ represents the category dimension, and $d_{e}$ denotes the embedding size. For the prediction layer, the time complexity of inner product is $\mathcal{O}(\sum_{l=1}^L |\mathcal{G}_u|d_l$), in which $|\mathcal{G}_u|$ is the number of items in the user-item bipartite graph. Assume that we have $m$ local updates per global update, the overall training complexity of MetaKG is $\mathcal{O}(m|\mathcal{G}_k| d_{l} d_{l-1}+|\mathcal{T}|d_{lstm}^2+|\mathcal{G}| d_{p} d_{e}+\sum_{l=1}^L |\mathcal{G}_u|d_l)$.

It is worth noting that traditional methods (i.e., KGAT) need to update all the parameters every epoch, which suffers from the slow convergence for its tremendous parameters. In contrast, MetaKG locally updates the aggregation layer only $m$ times ($m$ is usually small as validated in Section~\ref{sec:parameter analysis}), and globally updates only a few steps, which yields high performance. Hence, the overall training is efficient (to be reported in Section~\ref{sec:efficiency evaluation}).

\textbf{Parameters.} The model size largely depends on the knowledge graph embedding layer (e.g., nearly 11 million on Yelp2018 dataset), which is the same as the state-of-the-art method NFM~\cite{sigir17NFM}. In contrast, the weights of propagation process are much smaller (e.g., 15 thousand of three propagation layer on Yelp2018 dataset). It is worth mentioning that our two key components (i.e., collaborative-aware and knowledge-aware meta learners) do not need more parameters to achieve good generalization, thanks to the meta-learning framework.

Overall, MetaKG is a lightweight recommendation framework with high efficiency and adaptability, which is also demonstrated by empirically comparing MetaKG with various state-of-the-art methods in Section \ref{sec:efficiency evaluation}.



%% file: experiment.tex
\section{Experiments}
\label{sec:experiment}
In this section, we first present experimental settings, and then provide extensive experiments to evaluate our proposed MetaKG in terms of the recommendation performance, the model ablation, the model efficiency, the model scalability, and the parameter analysis, respectively.

\subsection{Experimental Settings}\label{sec:Experimental settings}
\textbf{Datasets.} To evaluate the performance of MetaKG, we use three public and popular benchmark datasets involving different scenarios: Amazon-book, Last-FM, and Yelp2018. These datasets are also adopted in the state-of-the-art methods~\cite{kdd19kgat}. For each dataset, $k$-core setting is used to ensure the quality of the dataset, meaning that we only keep users and items with at least $k$ interactions. Here, we use 10-score setting for all the three datasets. The detailed dataset descriptions are as follows.
\begin{itemize}[leftmargin=*]
\item \textbf{Yelp2018.\footnote{\url{https://www.yelp.com/dataset}}} This is a local business rating dataset collected by Yelp. We use the 2018 edition dataset of the Yelp challenge, and we take the local businesses like restaurants and bars as the items.
\item \textbf{Last-FM.\footnote{\url{https://grouplens.org/datasets/hetrec-2011}}} This is a music listening dataset provided by Last.fm online music system, where tracks are identified as items. We follow the work~\cite{kdd19kgat,www20KGPolicy} to take the subset of the dataset where the timestamp is from Jan, 2015 to June, 2015.
\item \textbf{Amazon-book.\footnote{\url{ http://jmcauley.ucsd.edu/data/amazon}}} We select Amazon-book from Amazon-review, a widely used dataset for product recommendation~\cite{www16amazon}. Here, if a user $u$ rates an item $i$, we set it as an observed interaction, which means $l_{ui}= 1$ in the user-item bipartite graph $\mathcal{G}_u$.
\end{itemize}

In addition, we derive the knowledge graphs from Freebase (for Amazon-book and Last-FM datasets), and the local business information network (for Yelp2018 datasets), which are widely used in preivous studies~\cite{kdd19kgat,sigir20ckan}. The overall statistics of datasets are summarized in Table~\ref{tab:data}.

\begin{table}[!]
\vspace{-2mm}
\caption{Statistics of the Datasets Used in Our Experiments}
\vspace{-3mm}
\label{tab:data}
\begin{tabular}{cll|l|l}
\hline
\multicolumn{2}{l|}{} & Yelp2018 & Last-FM & Amazon-book \\ \hline
\multirow{3}{*}{$\mathcal{G}_u$} & \multicolumn{1}{l|}{\#Users} & 45, 919 & 23, 566 & 70, 679 \\
 & \multicolumn{1}{l|}{\#Items} & 45, 538 & 48, 123 & 24, 915 \\
 & \multicolumn{1}{l|}{\#Interactions} & 1, 185, 068 & 3, 034, 796 & 847, 733 \\ \hline
\multirow{3}{*}{$\mathcal{G}_k$} & \multicolumn{1}{l|}{\#Entities} & 90, 961 & 58, 266 & 88, 572 \\
 & \multicolumn{1}{l|}{\#Relations} & 42 & 9 & 39 \\
 & \multicolumn{1}{l|}{\#Triplets} & 1, 853, 704 & 464, 567 & 2, 557, 746 \\ \hline
\end{tabular}
\vspace{-4mm}
\end{table}

\textbf{Cold-start Scenarios.} We follow the previous work~\cite{kdd20metaHIN} to construct the cold-start scenarios. Specifically, we first divide users and items into two groups (i.e., old ones and new ones) according to user's joining time (or user's first action time) and item's releasing time. Then, we split each dataset into training and testing data sets. For training data set, we only collect old user ratings for old items. For testing data, users and items are partitioned into three scenarios: User Cold-start (UC), (i.e. recommending old items for new users); Item Cold-start (IC), (i.e., recommending new items for old users); and User-Item Cold-start (UIC), (i.e., recommending new items for new users).

\textbf{Meta-learning Settings.} We collect the support and query sets following the previous work~\cite{kdd19melu}. Among the items rated by user $u$, we randomly select 10 items as the query set (\textit{i.e.} $|\mathcal{Q}_u | = 10$), and use the remaining items as the support set. We will also show the impact of the support set size on the model performance in Section~\ref{sec:parameter analysis}.

\textbf{Evaluation Metrics.} We treat all the items that user $u$ has not interacted with as the negative items. Then, each method outputs the user’s preference scores over all the items, except for the rated ones in the training set. We adopt two widely-used evaluation protocols~\cite{kdd19kgat, aaai19explainable, www20KGPolicy}, i.e., Recall@$K$ and normalized discounted cumulative gain at rank $K$ (NDCG@$K$). Here, we set $K$ to 20.

\textbf{Compared Methods.} We comprehensively compare our proposed MetaKG with traditional methods (FM and NFM), the embedding-based methods (CKE, CFKG and UGRec), the path-based methods (MCRec and RippleNet), the propagation-based methods (KGCN, KGAT and CKAN), and the state-of-the-art meta-learning method (MeLU and MetaHIN). Note that, we omit some methods (e.g., PER~\cite{wsdm14per} and DKN~\cite{www18dkn}), because they have been obviously defeated by recently proposed methods: KGCN~\cite{www19kgcn} and KGAT~\cite{kdd19kgat}.

\begin{itemize}[leftmargin=*]
	\item \textbf{FM}~\cite{sigir11FM} is a factorization benchmark, which considers the second-order interactions between users and items.
	\item \textbf{NFM}~\cite{sigir17NFM} is the state-of-the-art factorization method which uses MLP to learn nonlinear and high-order interaction signals.
	\item \textbf{CKE}~\cite{kdd16cke} is an embedding-based method, which applies TransR~\cite{aaai15transR} and CF to combine structural, visual, and textual knowledge in a unified framework.
	\item \textbf{CFKG}~\cite{alg18cfkg} applies TransE\cite{nips13transE} on the unified graph including users, items, entities, and relations, casting the recommendation task as the prediction of $(u, r, v)$ triplets.
	\item \textbf{UGRec}~\cite{sigir21UGRec} is the state-of-the-art embedding-based method which not only models directed knowledge-aware relations with TransD~\cite{acl15transD}, but also utilizes attentive mechanism to model undirected co-occurrence relations among items.
	\item \textbf{MCRec}~\cite{kdd18metapath} is a path-based model, which extracts qualified meta-paths as connectivities between users and items.
	\item \textbf{RipppleNet}~\cite{cikm18ripple} is the state-of-the-art path-based models, which uses a memory-like network, and propagates users’ potential preferences to enrich user representations.
	\item \textbf{KGCN}~\cite{www19kgcn} is one of the state-of-the-art propagation-based models, which extends GCN to the KG by selectively aggregating neighborhood information, to learn both structure and semantic information of the KG as well as users’ personalized interests.
	\item \textbf{KGAT}~\cite{kdd19kgat} is also one of the stat-of-the-art propagation-based models, which utilizes an attention mechanism to discriminate the importance of neighbors in CKG.
    \item \textbf{CKAN}~\cite{sigir20ckan} is based on KGAT, which utilizes different aggregation schemes on the user-item graph and KG, to encode knowledge association and collaborative signals.
    \item \textbf{MeLU}~\cite{kdd19melu} is a meta-learning method that only leverages interactions to learn embeddings of new users and items.
	\item \textbf{MetaHIN}~\cite{kdd20metaHIN} is the stat-of-the-art meta-learning method that captures multifaceted semantics on heterogeneous information networks to address cold-start problems.
\end{itemize}

\begin{table*}[tb!]
    \setlength{\tabcolsep}{1.4mm}{
    \centering
    \vspace{-7mm}
    \caption{Performance Comparison of Four Recommendation Scenarios on Three Datasets (“RI” indicates the relative improvement of our MetaKG over the corresponding baseline)}
    \vspace{-3mm}
    \label{tab:results}
    \renewcommand\arraystretch{1.3}
    \begin{tabular}{c||c||c|c|c||c|c|c||c|c|c}\hline
    \multirow{2}{*}{Scenario} &  \multirow{2}{*}{Methods} & \multicolumn{3}{c|}{Yelp2018} & \multicolumn{3}{c|}{Last-FM} & \multicolumn{3}{c}{Amazon-book} \\ \cline{3-11}
    &  & \multicolumn{1}{c|}{Recall@20} & \multicolumn{1}{c|}{NDCG@20} & \multicolumn{1}{c|}{RI} & \multicolumn{1}{c|}{Recall@20} & \multicolumn{1}{c|}{NDCG@20} & \multicolumn{1}{c|}{RI} & \multicolumn{1}{c|}{Recall@20} & \multicolumn{1}{c|}{NDCG@20} & \multicolumn{1}{c}{RI}\\ \hline
    \multirow{12}{*}{\begin{tabular}[c]{@{}c@{}}User Cold-start (UC)\\ Old items \\ for new users \end{tabular}}
     & FM & 0.0978 & 0.0812 & 45.74\% & 0.1917 & 0.1659 & 110.86\% & 0.1597 & 0.1420 & 116.17\% \\
     & NFM & 0.1174	& 0.0998 & 19.93\% & 0.2143	& 0.1905 & 86.06\%	& 0.1791 & 0.1573 & 94.00\% \\ \cline{2-11}
     & CKE & 0.1188	& 0.1050 & 16.18\% & 0.2632 & 0.2521 & 45.96\% & 0.2401 & 0.2391 & 35.72\% \\
     & CFKG & 0.0936 & 0.0811 & 48.95\%  & 0.2525 & 0.2362 & 53.92\% & 0.2180 & 0.2050 & 53.84\% \\
     & UGRec & 0.0711 & 0.0624 & 94.81\%	& 0.1912 & 0.1956 & 94.66\%	& 0.0778 & 0.0717 & 335.56\% \\  \cline{2-11}
     & MCRec & 0.0920 & 0.0793 & 51.95\%	& 0.1580	& 0.1532 & 141.70\%	& 0.1529 & 0.1641 & 105.51\% \\
     & RippleNet & 0.1001 & 0.0833 & 42.22\% & 0.1760 & 0.1607 & 123.51\%	& 0.1815 & 0.1741 & 82.93\% \\ \cline{2-11}
     & KGCN & 0.1123 & 0.0901 & 29.23\%	& 0.2206 & 0.2067 & 76.04\%	& 0.2445 & 0.2279 & 37.79\% \\
     & KGAT & 0.1240	& 0.1110	& 10.59\% & 0.2821 & 0.2744 & 35.16\%	& 0.2804 & 0.2712 & 17.92\% \\
     & CKAN & 0.0844 & 0.0802 & 57.74\%	& 0.1634 & 0.1649 & 129.26\%	& 0.2650 & 0.2567 & 24.67\% \\ \cline{2-11}
     & MeLU & 0.1102 & 0.1051 & 20.59\% & 0.2439 & 0.2344 & 57.25\% & 0.2265 & 0.2207 & 45.44\% \\
     & MetaHIN & 0.1260 & 0.1204 & 5.38\% & 0.2610 & 0.2628 & 43.68\% & 0.2569 & 0.2504 & 28.21\%      \\  \cline{2-11}
     & \textbf{MetaKG} & \textbf{0.1362} & \textbf{0.1236} & -- & \textbf{0.3927} & \textbf{0.3598} & -- & \textbf{0.3291} & \textbf{0.3213} & -- \\ \hline \hline

     \multirow{12}{*}{\begin{tabular}[c]{@{}c@{}}Item Cold-start (IC) \\ New items \\ for old users \end{tabular}}
     & FM & 0.0888 & 0.0755 & 78.25\% & 0.3574 & 0.3234 & 38.23\% & 0.2055 & 0.1792 & 16.57\% \\
     & NFM & 0.0844 & 0.0740 & 84.69\% & 0.3582 & 0.3192 & 39.02\% & 0.2084 & 0.1829 & 14.57\% \\ \cline{2-11}
     & CKE & 0.0888 & 0.0809 & 72.26\% & 0.3476 & 0.3112 & 42.92\% & 0.2167 & 0.1931 & 9.33\% \\
     & CFKG & 0.0917 & 0.0775 & 73.14\% & 0.3523 & 0.3259 & 38.65\% & 0.2073 & 0.1809 & 15.51\% \\
     & UGRec & 0.1349 & 0.1188 & 15.29\%	& 0.2623 & 0.2334 & 90.00\%	& 0.1716 & 0.1519 & 38.53\% \\  \cline{2-11}
     & MCRec & 0.0750 & 0.0691 & 102.85\% & 0.3605 & 0.3312 & 35.98\% & 0.2070 & 0.1573 & 24.47\% \\
     & RippleNet & 0.0920 & 0.0737 & 77.37\% & 0.3090 & 0.2777 & 60.45\% & 0.1710 & 0.1492 & 40.05\% \\ \cline{2-11}
     & KGCN & 0.0904 & 0.0718 & 81.32\% & 0.3270 & 0.3111 & 47.27\% & 0.1748 & 0.1529 & 36.83\% \\
     & KGAT & 0.0905 & 0.0788 & 72.83\% & 0.3445 & 0.3272 & 39.91\% & 0.2143 & 0.1902 & 10.78\% \\
     & CKAN & 0.1360 & 0.1204 & 14.06\%	& 0.2758 & 0.2946 & 65.00\%	& 0.1762 & 0.1693 & 29.51\% \\ \cline{2-11}
     & MeLU & 0.0712 & 0.0620 & 119.68\% & 0.2217 & 0.2119 & 116.72\% & 0.1920 & 0.1869 & 18.11\% \\
     & MetaHIN & 0.0670 & 0.0528 & 145.64\% & 0.2399 & 0.2266 & 101.47\% & 0.1990 & 0.1705 & 21.47\% \\ \cline{2-11}
     & \textbf{MetaKG} & \textbf{0.1571} & \textbf{0.1356} & -- & \textbf{0.4780} & \textbf{0.4616} & -- & \textbf{0.2336} & \textbf{0.2141} & -- \\ \hline \hline

     \multirow{12}{*}{\begin{tabular}[c]{@{}c@{}}User-Item \\ Cold-start (UIC) \\ New items \\ for new users \end{tabular}}
     & FM & 0.1181 & 0.0949 & 53.67\% & 0.3367 & 0.3043 & 33.21\% & 0.2412 & 0.1825 & 30.10\% \\
     & NFM & 0.1121 & 0.0982 & 54.95\% & 0.3390 & 0.3020 & 33.32\% & 0.2721 & 0.2265 & 9.60\% \\ \cline{2-11}
     & CKE & 0.1293 & 0.1156 & 32.99\% & 0.3118 & 0.2976 & 39.81\% & 0.2691 & 0.2206 & 11.71\% \\
     & CFKG & 0.0992 & 0.0819 & 80.42\% & 0.3208 & 0.3013 & 37.02\% & 0.2444 & 0.2102 & 19.98\% \\
     & UGRec & 0.1173 & 0.1023 & 48.41\%	& 0.2571 & 0.2369 & 72.68\%	& 0.2535 & 0.2216 & 14.72\% \\  \cline{2-11}
     & MCRec & 0.0910 & 0.0859 & 84.05\% & 0.1990 & 0.1857 & 121.63\% & 0.1800 & 0.1634 & 58.52\% \\
     & RippleNet & 0.1150 & 0.1025 & 49.76\% & 0.2150 & 0.2110 & 99.89\% & 0.1969 & 0.1435 & 62.68\% \\ \cline{2-11}
     & KGCN & 0.1189 & 0.1004 & 48.81\% & 0.3473 & 0.3060 & 30.90\% & 0.2653 & 0.2298 & 10.13\% \\
     & KGAT & 0.1498 & 0.1360 & 13.93\% & 0.3352 & 0.3182 & 30.41\% & 0.2670 & 0.2305 & 9.62\% \\
     & CKAN & 0.1462 & 0.1166 & 24.62\%	& 0.2854 & 0.2943 & 46.88\%	& 0.2750 & 0.2273 & 8.85\% \\ \cline{2-11}
     & MeLU & 0.1308 & 0.1237 & 27.94\% & 0.2071 & 0.1875 & 116.37\% & 0.2498 & 0.2014 & 21.41\% \\
     & MetaHIN & 0.1430 & 0.1382 & 15.82\% & 0.2030 & 0.1871 & 118.67\% & 0.2656 & 0.2285 & 10.39\% \\ \cline{2-11}
     & \textbf{MetaKG} & \textbf{0.1747} & \textbf{0.1513} & -- & \textbf{0.4227} & \textbf{0.4287} & -- & \textbf{0.2857} & \textbf{0.2587} & -- \\ \hline \hline

    \multirow{12}{*}{\begin{tabular}[c]{@{}c@{}}Non-cold-start \\ Old items \\ for old users \end{tabular}}
     & FM & 0.0916 & 0.0779 & 77.43\% & 0.2193 & 0.1948 & 103.42\% & 0.1168 & 0.1073 & 111.06\% \\
     & NFM & 0.1002 & 0.0870 & 60.47\% & 0.2677 & 0.2344 & 67.91\% & 0.1368 & 0.1256 & 80.25\% \\ \cline{2-11}
     & CKE & 0.1353 & 0.1214 & 16.86\% & 0.2952 & 0.2804 & 45.95\% & 0.2233 & 0.2181 & 6.95\% \\
     & CFKG & 0.0802 & 0.0693 & 100.98\% & 0.2850 & 0.2625 & 53.59\% & 0.1239 & 0.1162 & 96.82\% \\
     & UGRec & 0.0592 & 0.0499 & 175.81\%	& 0.1898 & 0.1927 & 119.55\%	& 0.0398 & 0.0351 & 532.93\% \\  \cline{2-11}
     & MCRec & 0.0670 & 0.0529 & 152.29\% & 0.1220 & 0.1078 & 266.68\% & 0.0670 & 0.0619 & 266.84\% \\
     & RippleNet & 0.0690 & 0.0631 & 126.97\% & 0.1910 & 0.1721 & 131.82\% & 0.1061 & 0.0864 & 147.96\% \\ \cline{2-11}
     & KGCN & 0.0645 & 0.0579 & 145.09\% & 0.2006 & 0.1882 & 116.15\% & 0.1117 & 0.0910 & 135.47\% \\
     & KGAT & 0.1270 & 0.1166 & 23.07\% & 0.3004 & 0.2892 & 42.45\% & 0.2221 & 0.2188 & 7.06\% \\
     & CKAN & 0.0868 & 0.0692 & 93.73\%	& 0.2071 & 0.2042 & 104.16\%	& 0.1040 & 0.0960 & 136.44\% \\ \cline{2-11}
     & MeLU & 0.1266 & 0.1183 & 22.38\% & 0.2828 & 0.2755 & 50.42\% & 0.1709 & 0.1630 & 41.46\% \\
     & MetaHIN & 0.1420 & 0.1226 & 13.56\% & 0.3029 & 0.2886 & 42.02\% & 0.2160 & 0.2181 & 8.72\% \\ \cline{2-11}
     & \textbf{MetaKG} & \textbf{0.1562} & \textbf{0.1436} & -- & \textbf{0.4221} & \textbf{0.4176} & -- & \textbf{0.2344} & \textbf{0.2376} & -- \\ \hline
    \end{tabular}
    \vspace{-5mm}
    }
\end{table*}

\textbf{Parameters Settings.} We apply grid search for hyper-parameter tuning. We adopt Adam to optimize MetaKG, and utilize the default Xavier initializer~\cite{ICML10Xavier} to initialize the parameters. The embedding size is fixed to 64 for users, entities and relations. For all datasets, we set the global learning rate to 0.0001, local learning rate to 0.01 and scheduler learning rate to 0.001 after tuning amongst \{0.1, 0.01, ..., 0.0001\}. For adaptive scheduler module, the bidirectional LSTM with hidden units set to 10 is used for embedding both the loss and the gradient similarity. To explore the high-order connectivity information, we stack three propagation layers with three fusion gates. Finally, we implement MetaKG in Python and Pytorch\footnote{\url{ https://pytorch.org}}. All experiments were conducted on a server with Intel(R) Xeon(R) Silver 4216, 2.10GHz CPU, 256GB RAM, and a NVIDIA A100 GPU.
The used datasets and source codes of MetaKG are available online\footnote{\url{ https://github.com/ZJU-DBL/MetaKG}}.

For baselines, we set their parameters under the guidance of literature. The user, item, entity and relation embedding dimensions are fixed to 64 for all baselines. The learning rate is tuned amongst \{0.0001, 0.001, 0.01\}, and the coefficient of $L_{2}$ normalization is searched in \{$10^{-6}$, $10^{-5}$ , ..., $10^{-1}$\}. For NFM, we set the MLP layers size to [64, 32, 1] with dropout ratio 0.1. For CKE, CFKG, and KGAT, the KG training batch size is set to 2048, and the learning rate is 0.0001. Specifically, for KGAT, we set the depth to three with the hidden size [64, 32, 16], and set dropout ratio to 0.1. For RippleNet, we set the number of hops and the memory size to 2 and 8, respectively. For UGRec, the margin value for hinge loss is set to 1.5. For CKAN, we set the size of triple set for user and item to 16 and 64 respectively. For MeLU and MetaHIN,  both of the local and global learning rate are set to 0.0001, and a single local update is applied to this model.

\subsection{Performance Comparison}\label{sec:performance}
In this section, we investigate the performance of MetaKG compared with state-of-the-art methods under three cold-start scenarios and one non-cold-start scenario, using three public datasets. Table~\ref{tab:results} summarizes the results.

\textbf{Cold-start Scenarios.} As observed in Table~\ref{tab:results}, the results of three cold-start scenarios (i.e., UC, IC, and UIC) show that our MetaKG consistently yields the best performance among all the methods on all the datasets. Specifically, MetaKG achieves 62.14\%, 74.95\%, and 67.28\% of average improvement over the competitors in Yelp2018, Last-FM, and Amazon-book, respectively. The second observation is that embedding-based methods (i.e., CKE, CFKG and UGRec) perform better than traditional methods (i.e., FM and NFM) in most cases, indicating that the effectiveness of entity embeddings for recommendation. The third observation is that the path-based methods (i.e., MCRec and RippleNet) exhibit poor competitiveness compared with the embedding-based methods. This is because the handcraft meta-paths highly depend on the massive user-item interactions and thus are insufficient to model implicit relations in cold-start scenarios. Moreover, the propagation-based methods (i.e., KGCN, KGAT and CKAN) perform well among all the baselines since they enable to capture collaborative signal in an end-to-end manner, while the meta-learning based methods (i.e., MetaHIN and our MetaKG) achieve better performance than the propagation-based methods. This further demonstrates the effectiveness of meta-learning for cold-start recommendation problems. However, MetaHIN still under-performs our MetaKG in all scenarios. The reason is that MetaHIN only considers preference learning with handcraft meta-paths, while MetaKG automatically explores the high-order connectivity within each task, and captures entities' general knowledge to further improve the recommendation quality across tasks.

\textbf{Non-cold-start Scenario.} We also evaluate the performance of MetaKG considering non-cold-start scenario, as depicted in Table~\ref{tab:results}. As expected, our MetaKG still achieves the best performance among all methods, which further confirms the robustness of MetaKG. The reason behind is that the evaluated datasets are very sparse (e.g., the sparsity of Yelp 2018 is 99.93\%), and our method could alleviate the sparsity problems and have good generalization, thanks to the novel design of two meta-learners. Besides, it is worth mentioning that the performance of competitors are unstable among different scenarios. In contrast, our MetaKG model achieves the state-of-the-art recommendation performance considering both cold-start scenarios and non-cold-start scenario.

\subsection{Ablation Study}\label{sec:ablation study}

\begin{figure}[tb!]
	\centering
	\includegraphics[width=0.35\textwidth]{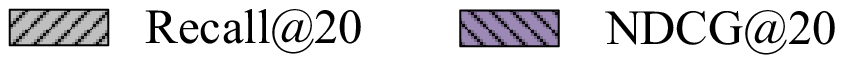}\\
	\vspace{-1mm}
	\subfigure[Yelp2018 dataset]{
		\includegraphics[width=0.233\textwidth]{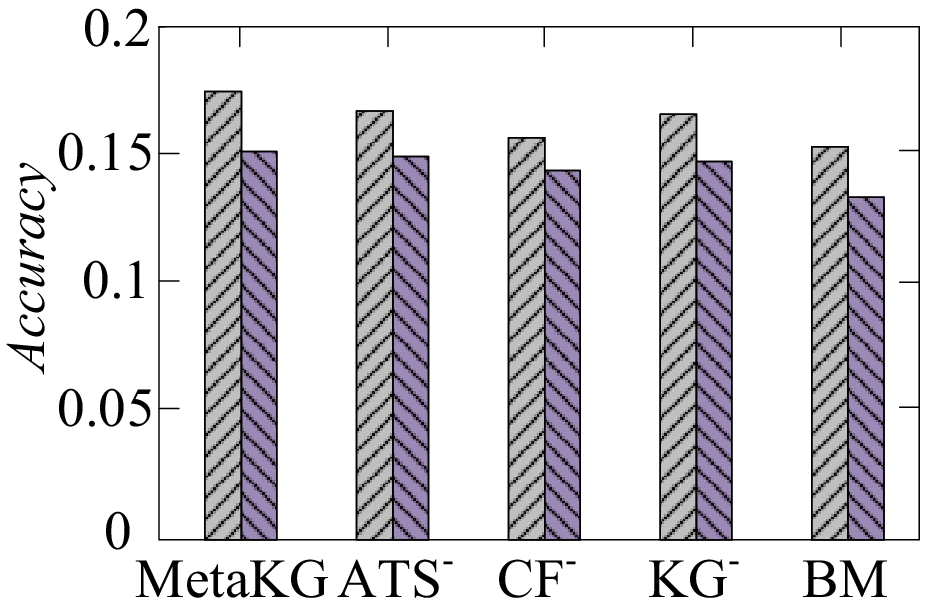}}
	\hspace{1.4mm}
	\hspace{-0.25cm}
	\subfigure[Last-FM dataset]{
		\includegraphics[width=0.233\textwidth]{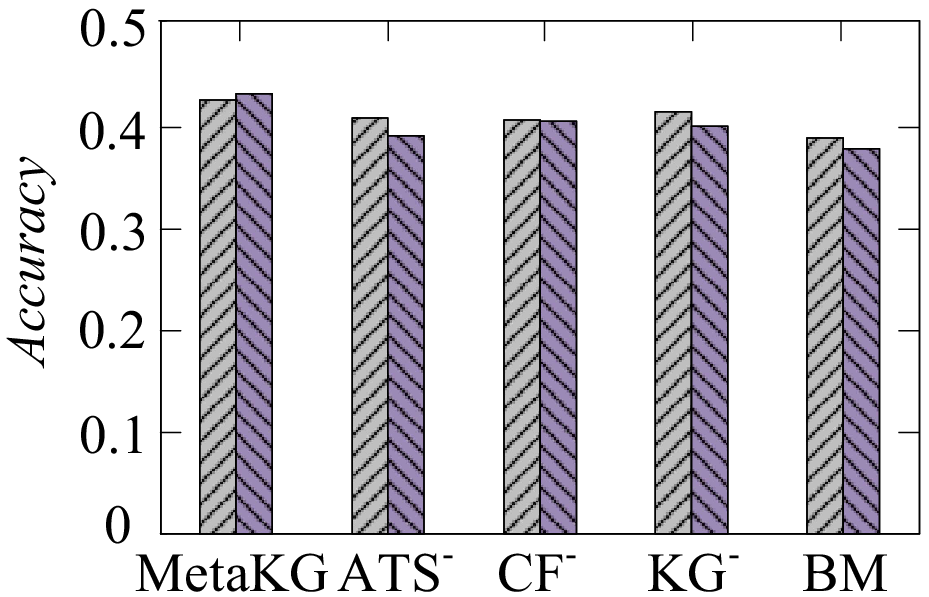}}
	\vspace{-5mm}
	\caption{Ablation Study of MetaKG in UIC Scenario}
	\label{fig:ablation}
	\vspace{-2mm}
\end{figure}

In this subsection, we provide the insights into the effectiveness of each component in MetaKG. The results are shown in Fig.~\ref{fig:ablation}, where (i) ``MetaKG'' represents its completed model; (ii) ``ATS$^-$'' denotes MetaKG without adaptive task scheduler; (iii) ``CF$^-$'' denotes MetaKG without colloborative-aware meta learner; (iv) ``KG$^-$'' represents MetaKG without knowledge-aware meta learner; and (v) ``BM'' stands for the base model of MetaKG, which is trained without the meta-learning framework. Specifically, we optimize all the learnable parameters (i.e., $\theta$, $\phi$ and $\omega$) with BPR loss function, and directly use all training data to update them without the local update scheme. The details of base model can be refereed in Section~\ref{sec:base model}.

\textbf{MetaKG vs. ATS$^{-}$.} Firstly, we remove the adaptive task scheduler module to verify its effectiveness. As observed, Recall@20 drops 3.45\%, and NDCG@20 drops 5.07\% on the Last-FM dataset, after sampling tasks uniformly as usual. Similar observations can also be found on the Yelp2018 and Amazon-book. This shows that adaptive task scheduler could effectively select informative tasks and filter out noisy tasks for better meta learners training.

\textbf{MetaKG vs. CF$^{-}$.} To verify the effectiveness of the collaborative-aware meta learner, we remove it by freezing the parameteres $\gamma$ in the aggregation layer within tasks. As observed, taking Last-FM as an example, MetaKG improves the Recall@20 from 0.4096 to 0.4227, and boosts NDCG@20 from 0.4081 to 0.4287, compared with CF$^{-}$. Similar observations can also be found on the Yelp2018 and Amazon-book datasets. This shows that the collaborative-aware meta leaner can effectively capture \textit{meta} high-order user-item relations to achieve better recommendation performance.

\textbf{MetaKG vs. KG$^{-}$.} The major difference between MetaKG and existing meta-learning models lies in whether multi-objective optimization is utilized to enrich semantic embeddings of entities in KG. By removing the knowledge-aware meta learner, we only leverage the initial embeddings to predict the user preference, without optimizing the KG loss. As observed, Recall@20 drops 5.98\%, and NDCG@20 drops 6.57\%, after removing the knowledge-aware meta learner from MetaKG. This further proves that the knowledge-aware meta learner benefits the recommendation performance improvement.

\textbf{MetaKG vs. BM.} Finally, we explore the performance of MetaKG without meta-learning guiding, meaning that we only adopt the base model for cold-start recommendations. Compared with BM, MetaKG achieves significant improvement across all the metrics and datasets. The reason of the dramatic performance degradation is that the base model is trained without the carefully designed meta learners, and thus, it easily incurs overfitting when the data is scarce, and lacks the fast adaptation ability for new users or new items, as discussed in Section~\ref{sec:introduction}.

\subsection{Efficiency Evaluation}\label{sec:efficiency evaluation}
Next, we inspect the model efficiency in terms of both model training (denoted as \textbf{T}) and inference (denoted as \textbf{I}) phases. Due to the limited space, we only compare MetaKG with the most popular traditional method (i.e., NFM), the state-of-the-art KG-based method (i.e., KGAT), and the state-of-the-art meta-learning method (i.e., MetaHIN), as depicted in Table~\ref{tab:efficiency}. In addition, we only study the UIC scenario, as other scenarios show similar performance. Overall, MetaKG achieves high efficiency for both training and inference phases. Specifically, in terms of the training phase, MetaKG finishes each epoch within 3.2 seconds on Yelp2018, which is faster than the best competitor (i.e., NFM). In terms of the inference phase, MetaKG also performs well (i.e., only 0.3 seconds per epoch on all the datasets). This is because our MetaKG only needs a few gradients steps to gain good generalizations with the benefits of meta-learning, while the other methods need much more tune to get satisfying results. Note that, although MetaHIN is a meta-learning based approach, it is also inferior to MetaKG. This is because MetaHIN needs time-consuming process to construct semantic meta-paths, which is time-sensitive and not feasible for online recommendation.

\begin{table}[tb!]
\centering
\caption{Running Time (s/epoch) of Training and Inference}
\vspace{-3mm}
\setlength{\tabcolsep}{2.8mm}{
\begin{tabular}{c|cc|cc|cc}
\toprule
\multirow{2}{*}{\textbf{Methods}} & \multicolumn{2}{c|}{\textbf{Yelp2018}} & \multicolumn{2}{c|}{\textbf{Last-FM}} & \multicolumn{2}{c}{\textbf{Amazon-book}} \\ \cline{2-7}
  & T   & I  & T  & I & T  & I \\ \hline
NFM & 4.9 & 1.5 & 20.6   & 7.4 & 7.5 & 1.2  \\
KGAT & 107.2  & \textbf{0.3} & 51.2   & \textbf{0.3}   & 197.6  & \textbf{0.3}  \\
MetaHIN & 220.1 & 4.8 & 97.5   & 9.1   & 65.3  & 1.5  \\
MetaKG & \textbf{3.2} &\textbf{0.3} & \textbf{3.9}   & \textbf{0.3} & \textbf{3.8}  & \textbf{0.3}  \\ \bottomrule
\end{tabular}}
\label{tab:efficiency}
\vspace{-2mm}
\end{table}

\subsection{Scalability  Evaluation}\label{sec:scalability evaluation}
\begin{figure}[tb]
	\centering
	\vspace{2mm}
	\includegraphics[width=0.4\textwidth]{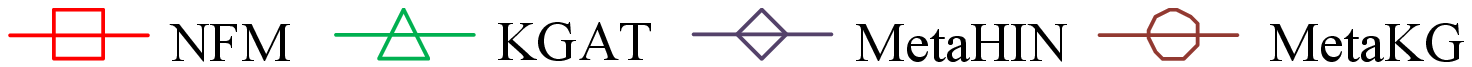}\\
	\subfigure[Yelp2018 dataset]{
		\includegraphics[width=0.233\textwidth]{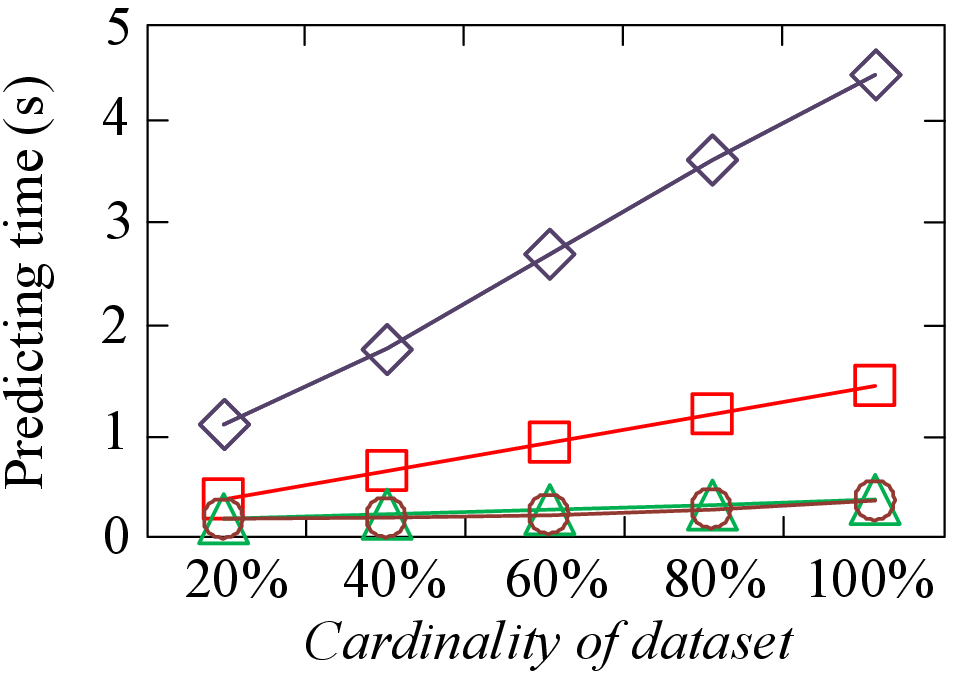}}
	\hspace{1.4mm}
	\hspace{-0.25cm}
	\subfigure[Last-FM datase]{
		\includegraphics[width=0.233\textwidth]{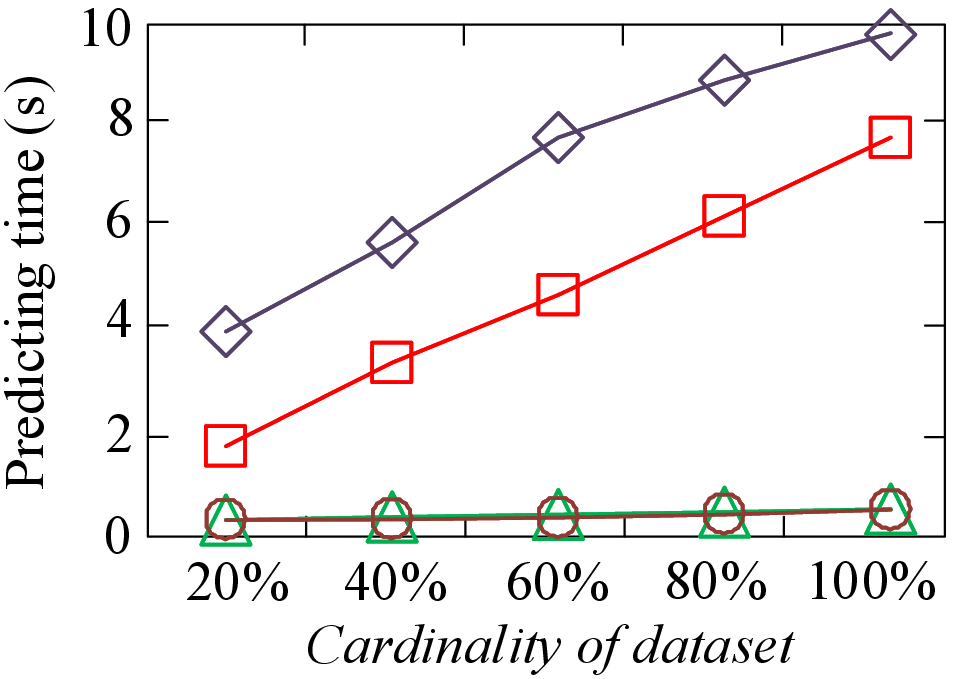}}
	\vspace{-5mm}
	\caption{Scalability Evaluation in UIC Scenario}
	\label{fig:scalability}
	\vspace{-3mm}
\end{figure}

We proceed to explore the scalability of MetaKG as it is important for online recommendation. We utilize three different kinds of state-of-the-art methods (i.e., NFM, KGAT and MetaHIN) as competitors, and vary the cardinality of dataset (i.e., the proportion of the test data size to the entire dataset) from 20\% to 100\%. Fig.~\ref{fig:scalability} plots the UIC scenario results on Yelp2018 and Last-FM datasets, while other scenarios are omitted due to similar trends. The first observation is that the predicting time increases with the cardinality. Second, MetaKG offers significant performance improvement over existing methods, which costs 0.3 seconds to predict the entire yelp2018 dataset. Finally, the performance of MetaKG is less affected by the growth of cardinality than other methods.
Note that, although KGAT achieves competitive performance in scalability, it fails to tackle cold-start scenarios as mentioned in Section~\ref{sec:performance}. Overall, MetaKG is capable of large-scale online recommendation.

\subsection{Parameter Analysis}\label{sec:parameter analysis}
Finally, we investigate the impact of parameters (i.e., the number of local updates, the support set size, and the embedding dimension) on the model performance. Here, we only utilize the UIC scenario due to the similar observations captured from other cold-start scenarios.

\begin{figure}[tb]
	\centering
	
	\includegraphics[width=0.35\textwidth]{experiment_fig/bar-title.eps}\\
	\vspace{-2mm}
	\subfigure[Yelp2018 dataset]{
		\includegraphics[width=0.233\textwidth]{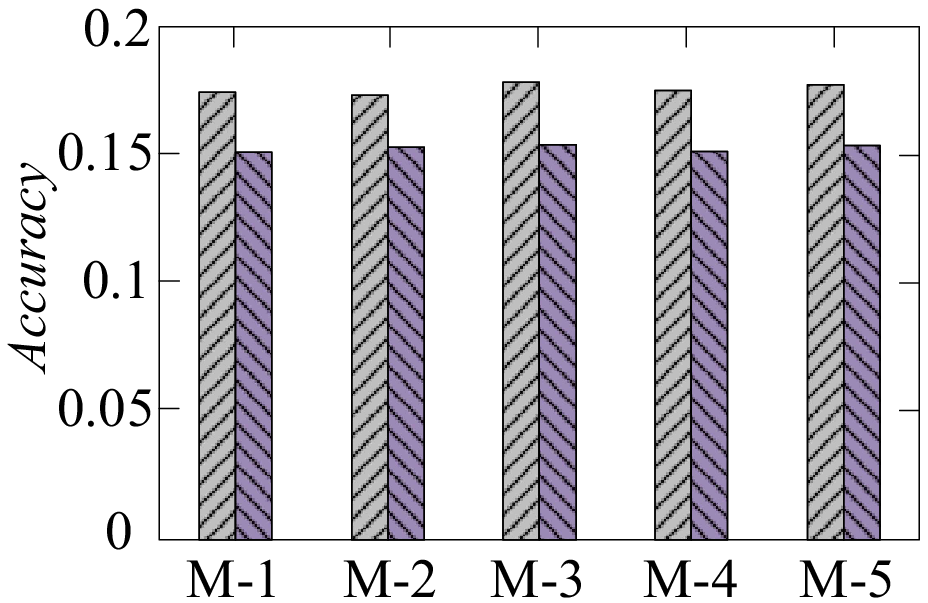}}
	\hspace{1.4mm}
	\hspace{-0.25cm}
	\subfigure[Last-FM dataset]{
		\includegraphics[width=0.233\textwidth]{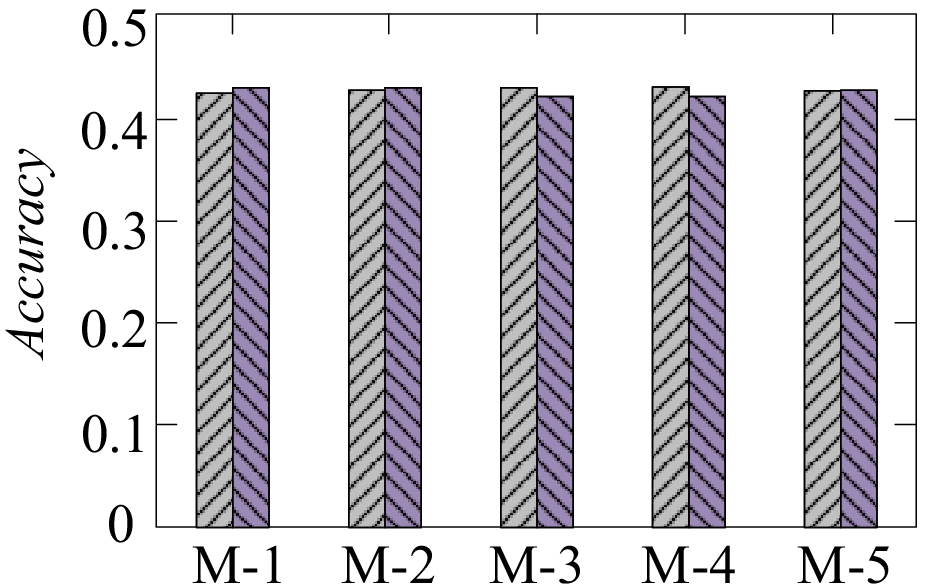}}
	\vspace{-5mm}
	\caption{Model Performance vs. the Number of Local Updates}
	\label{fig:localupdate}
	\vspace{-3mm}
\end{figure}

\textbf{Effect of Number of Local Updates.} We qualitatively analyze the effect of the local updates on the model performance, which leverages collaborative-aware meta learner to provide personalized preference information for recommendations. We denote the proposed model with $m$ local updates as M-$m$. Fig.~\ref{fig:localupdate} depicts the results by changing numbers of local updates on Yelp2018 and Last-FM datasets. As observed, MetaKG achieves remarkable improvement on all the datasets and metrics. In addition, the performance of MetaKG is stable with the increase of the number of local updates. Thus, our MetaKG model can be adapt quickly to user preference because a single local update is enough and sufficient, indicating that MetaKG shows high potential for online recommendations.

\begin{figure}[tb!]
	\centering
	\includegraphics[width=0.5\textwidth]{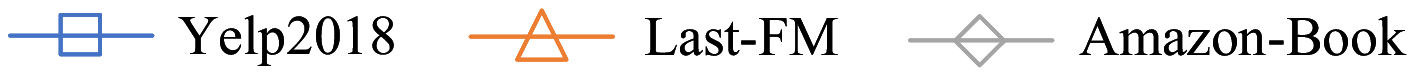}\\
	\vspace{-2mm}
	\subfigure[Performance at Recall@20]{\includegraphics[width=0.233\textwidth]{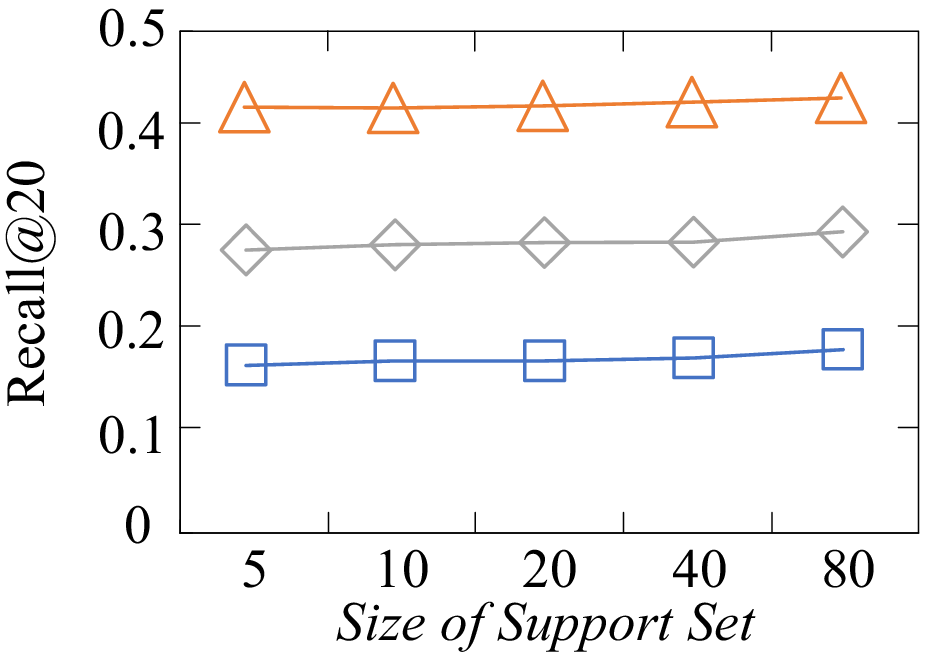}}
	\hspace{1.4mm}
	\hspace{-0.25cm}
	\subfigure[Performance at NDCG@20]{
		\includegraphics[width=0.233\textwidth]{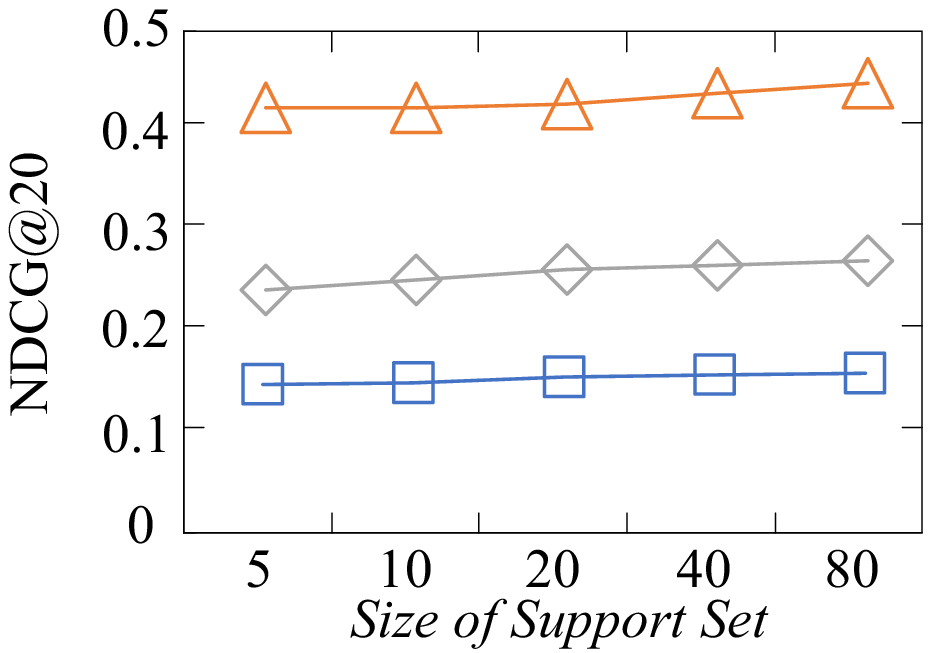}}
	\vspace{-2.5mm}
	\caption{Model Performance vs. Support Set Size}
	\label{fig:supportset}
	\vspace{-3mm}
\end{figure}

\textbf{Effect of Support Set Size.}
We verify the model performance when varying the size of support set (i.e., $|S_u|$) from 5 to 80. The corresponding results are shown in Fig.~\ref{fig:supportset}. Overall, the performance of MetaKG becomes better with the growth of the support set size (i.e., more training data). Nonetheless, the performance improves slowly as the size of support set ascends. This implies that MetaKG can achieve high performance via very few interactions/sampling data, which validates the robustness of MetaKG especially in cold-start scenarios.

\begin{figure}[tb!]
	\centering
	\includegraphics[width=0.5\textwidth]{experiment_fig/title.eps}\\
	\vspace{-2mm}
	\subfigure[Performance at Recall@20]{
		\includegraphics[width=0.233\textwidth]{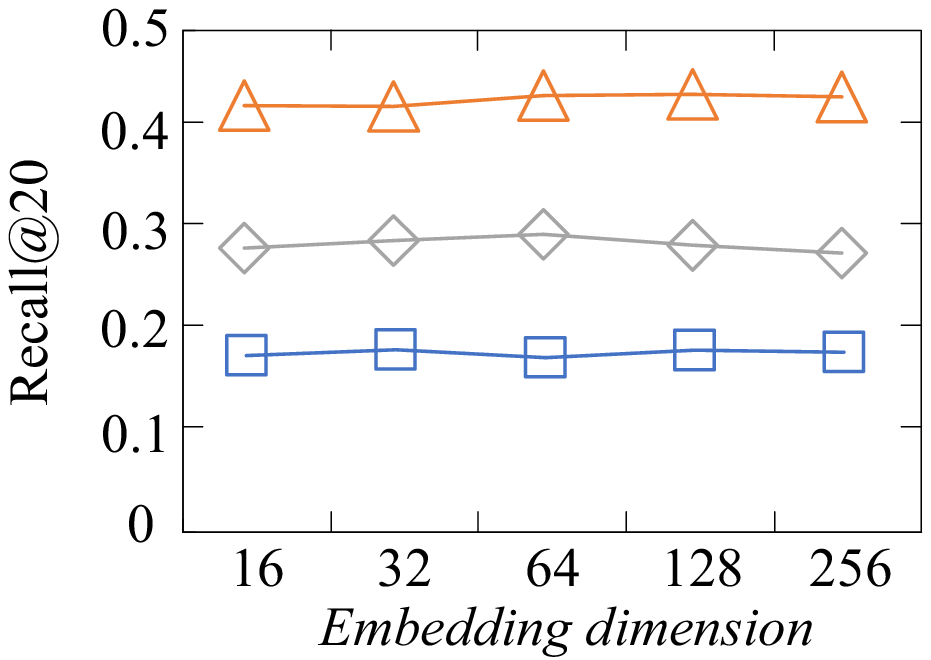}}
	\hspace{1.4mm}
	\hspace{-0.25cm}
	\subfigure[Performance at NDCG@20]{
		\includegraphics[width=0.233\textwidth]{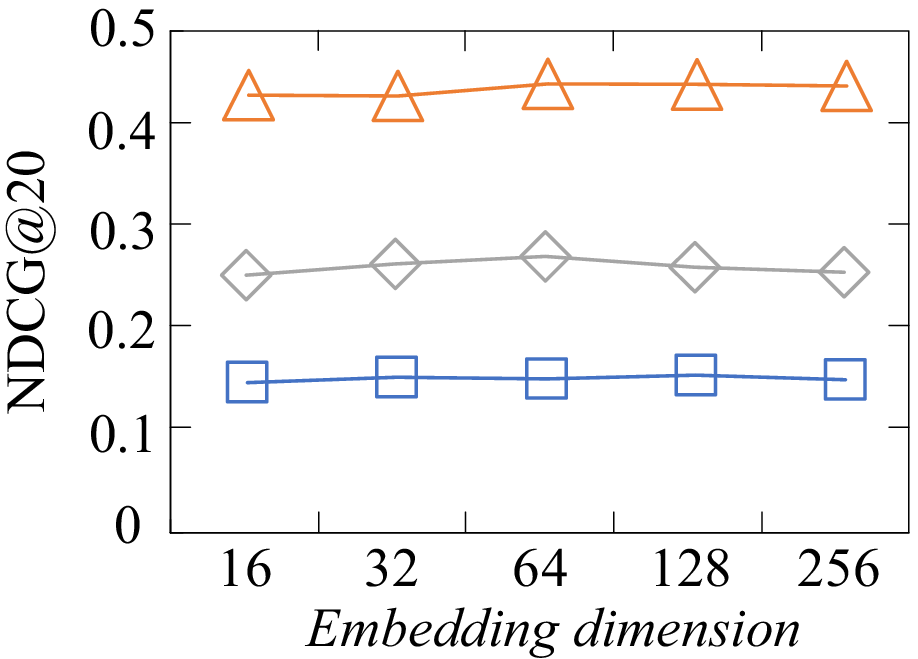}}
	\vspace{-5mm}
	\caption{Model Performance vs. Embedding Dimension}
	\label{fig:embedding}
	\vspace{-3mm}
\end{figure}

\textbf{Effect of Embedding Dimension.} Last but not the least, we explore how the embedding dimension $d_e$ affect the model performance. The results are illustrated in Fig.~\ref{fig:embedding}. As observed, MetaKG achieves the best performance when the dimension is 64 in most cases. Also, the performance is stable with the growth of $d_e$, which confirms the robustness of MetaKG w.r.t the embedding dimension.

%% file: conclusion.tex
\section{Conclusions}\label{sec:conclusions}

In this paper, we propose MetaKG, a meta-learning framework for cold-start recommendation on the knowledge graph, which consists of two meta learners, including the collaborative-aware meta learner and knowledge-aware meta learner, to achieve the generality within and across user preference learning tasks. Specifically, the collaborative-aware meta learner learns users' preference within each task, while the knowledge-aware meta learner enables to capture the semantic representations of KG shared across different tasks. Besides, we devise a novel adaptive task scheduler to select the informative task for meta learning. Extensive experiments on three real data sets demonstrate that MetaKG significantly outperforms all the state-of-the-art baselines with high efficiency and robustness. Overall, to the best of our knowledge, it is the first work to explore the meta-learning in KG-based recommendation, where we capture the prior knowledge for different learning process. In the future, we would like to explore interpretability in KG for explainable cold-start recommendations. 
In addition, combining our method with recent non-sampling learning methods is also attractive. 

%% file: acknowledgments.tex
\section*{Acknowledgments}

This work was supported by the NSFC under Grants No. (62025206, 61972338, and 62102351). Yunjun Gao is the corresponding author of the work.

%% file: main.bbl
\begin{thebibliography}{10}

\bibitem{MANN}
S.~Adam, B.~Sergey, B.~Matthew, W.~Daan, and L.~Timothy.
\newblock Meta-learning with memory-augmented neural networks.
\newblock In {\em ICML}, pages 1842--1850, 2016.

\bibitem{alg18cfkg}
Q.~Ai, V.~Azizi, X.~Chen, and Y.~Zhang.
\newblock Learning heterogeneous knowledge base embeddings for explainable
  recommendation.
\newblock {\em Algorithms}, 11(9):137--146, 2018.

\bibitem{nips13transE}
A.~Bordes, N.~Usunier, A.~García-Durán, J.~Weston, , and O.~Yakhnenko.
\newblock Translating embeddings for modeling multi-relational data.
\newblock In {\em NeurIPS}, pages 2787--2795, 2013.

\bibitem{nips17high_variance}
H.-S. Chang, E.~Learned-Miller, and A.~McCallum.
\newblock Active bias: Training more accurate neural networks by emphasizing
  high variance samples.
\newblock In {\em NeurIPS}, pages 1002--1012, 2017.

\bibitem{maml}
F.~Chelsea, A.~Pieter, and S.~Levine.
\newblock Model-agnostic meta-learning for fast adaptation of deep networks.
\newblock In {\em ICML}, pages 1126--1135, 2017.

\bibitem{dlrs16wide-deep}
H.-T. Cheng, L.~Koc, J.~Harmsen, T.~Shaked, T.~Chandra, H.~Aradhye,
  G.~Anderson, G.~Corrado, W.~Chai, M.~Ispir, R.~Anil, Z.~Haque, L.~Hong,
  V.~Jain, X.~Liu, and H.~Shah.
\newblock Wide \& deep learning for recommender systems.
\newblock In {\em Proceedings of the 1st Workshop on Deep Learning for
  Recommender Systems}, pages 7--10, 2016.

\bibitem{14gru}
J.~Chung, C.~Gulcehre, K.~Cho, and Y.~Bengio.
\newblock Empirical evaluation of gated recurrent neural networks on sequence
  modeling.
\newblock In {\em arXiv preprint arXiv:1412.3555}, 2014.

\bibitem{recsys19basepaper}
M.~F. Dacrema, P.~Cremonesi, and D.~Jannach.
\newblock Are we really making much progress? a worrying analysis of recent
  neural recommendation approaches.
\newblock In {\em RecSys}, pages 101--109, 2019.

\bibitem{kdd19metaLSTMRec}
Z.~Du, X.~Wang, H.~Yang, J.~Zhou, and J.~Tang.
\newblock Sequential scenario-specific meta learner for online recommendation.
\newblock In {\em KDD}, pages 2895--2904, 2019.

\bibitem{relationNN}
S.~Flood, Y.~Yongxin, Z.~Li, X.~Tao, T.~P. HS, and H.~T. M.
\newblock Learning to compare: Relation network for few-shot learning.
\newblock In {\em CVPR}, pages 1199--1208, 2018.

\bibitem{ICML10Xavier}
X.~Glorot and Y.~Bengio.
\newblock Understanding the difficulty of training deep feedforward neural
  networks.
\newblock In {\em AISTATS}, pages 249--256, 2010.

\bibitem{nips17graphSage}
W.~L. Hamilton, R.~Ying, and J.~Leskovec.
\newblock Inductive representation learning on large graphs.
\newblock In {\em NeurIPS}, pages 1025--1035, 2017.

\bibitem{www16amazon}
R.~He and J.~McAuley.
\newblock Ups and downs: Modeling the visual evolution of fashion trends with
  one-class collaborative filtering.
\newblock In {\em WWW}, pages 507--517, 2016.

\bibitem{sigir17NFM}
X.~He and T.-S. Chua.
\newblock Neural factorization machines for sparse predictive analytics.
\newblock In {\em SIGIR}, pages 355--364, 2017.

\bibitem{www17ncf}
X.~He, L.~Liao, H.~Zhang, L.~Nie, X.~Hu, and T.-S. Chua.
\newblock Neural collaborative filtering.
\newblock In {\em WWW}, pages 173--182, 2017.

\bibitem{97lstm}
S.~Hochreiter and J.~Schmidhuber.
\newblock Long short-term memory.
\newblock {\em Neural computation}, 9:1735--1780, 1997.

\bibitem{kdd18metapath}
B.~Hu, C.~Shi, W.~X. Zhao, and P.~S. Yu.
\newblock Leveraging meta-path based context for top- n recommendation with a
  neural co-attention model.
\newblock In {\em KDD}, pages 1531--1540, 2018.

\bibitem{icdm08cf}
Y.~Hu, Y.~Koren, and C.~Volinsky.
\newblock Collaborative filtering for implicit feedback datasets.
\newblock In {\em ICDM}, pages 263--272, 2008.

\bibitem{nips17proto}
S.~Jake, S.~Kevin, and R.~Zemel.
\newblock Prototypical networks for few-shot learning.
\newblock In {\em NeurIPS}, pages 4080–--4090, 2017.

\bibitem{acl15transD}
G.~Ji, S.~He, L.~Xu, K.~Liu, and J.~Zhao.
\newblock Knowledge graph embedding via dynamic mapping matrix.
\newblock In {\em ACL}, pages 687--696, 2015.

\bibitem{iclr17gcn}
T.~N. Kipf and M.~Welling.
\newblock Semi-supervised classification with graph convolutional networks.
\newblock In {\em ICLR}, 2017.

\bibitem{SiameseNN}
G.~Koch, R.~Zemel, and R.~Salakhutdinov.
\newblock Siamese neural networks for one-shot image recognition.
\newblock In {\em ICML deep learning workshop}, volume~2, 2015.

\bibitem{kdd08cf}
Y.~Koren.
\newblock Factorization meets the neighborhood: A multifaceted collaborative
  filtering model.
\newblock In {\em KDD}, pages 426--434, 2008.

\bibitem{kdd19melu}
H.~Lee, J.~Im, S.~Jang, H.~Cho, and S.~Chung.
\newblock {MeLU}: Meta-learned user preference estimator for cold-start
  recommendation.
\newblock In {\em KDD}, pages 1073--1082, 2019.

\bibitem{www10content-rec}
L.~Li, W.~Chu, J.~Langford, , and R.~E. Schapire.
\newblock A contextual- bandit approach to personalized news article
  recommendation.
\newblock In {\em WWW}, pages 661--670, 2010.

\bibitem{aaai15transR}
Y.~Lin, Z.~Liu, M.~Sun, Y.~Liu, and X.~Zhu.
\newblock Learning entity and relation embeddings for knowledge graph
  completion.
\newblock In {\em AAAI}, pages 2181--2187, 2015.

\bibitem{kdd20metaHIN}
Y.~Lu, Y.~Fang, and C.~Shi.
\newblock Meta-learning on heterogeneous information networks for cold-start
  recommendation.
\newblock In {\em KDD}, pages 1563--1573, 2020.

\bibitem{www20metaselector}
M.~Luo, F.~Chen, P.~Cheng, Z.~Dong, X.~He, J.~Feng, and Z.~Li.
\newblock Metaselector: Meta-learning for recommendation with user-level
  adaptive model selection.
\newblock In {\em WWW}, pages 2507--2513, 2020.

\bibitem{SNAIL}
N.~Mishra, M.~Rohaninejad, X.~Chen, and P.~Abbeel.
\newblock A simple neural attentive meta-learner.
\newblock In {\em ICLR}, 2018.

\bibitem{nips16matchnet}
V.~Oriol, B.~Charles, L.~Timothy, kavukcuoglu koray, and D.~Wierstra.
\newblock Matching networks for one shot learning.
\newblock In {\em NeurIPS}, pages 3630--3638, 2016.

\bibitem{sigir19ctr}
F.~Pan, S.~Li, X.~Ao, P.~Tang, and Q.~He.
\newblock Warm up cold-start advertisements: Improving ctr predictions via
  learning to learn id embeddings.
\newblock In {\em SIGIR}, pages 695--704, 2019.

\bibitem{iclr17opti}
S.~Ravi and H.~Larochelle.
\newblock Optimization as a model for few-shot learning.
\newblock In {\em ICLR}, 2017.

\bibitem{uai09BPRloss}
S.~Rendle, C.~Freudenthaler, Z.~Gantner, and L.~Schmidt-Thieme.
\newblock Bpr: Bayesian personalized ranking from implicit feedback.
\newblock In {\em UAI}, pages 452--461, 2009.

\bibitem{sigir11FM}
S.~Rendle, Z.~Gantner, C.~Freudenthaler, and L.~Schmidt-Thieme.
\newblock Fast context-aware recommendations with factorization machines.
\newblock In {\em SIGIR}, pages 635--644, 2011.

\bibitem{metalearning01}
V.~Ricardo and D.~Youssef.
\newblock A perspective view and survey of meta-learning.
\newblock {\em Artificial Intelligence Review}, pages 77--95, 2001.

\bibitem{cikm20mkgat}
R.~Sun, X.~Cao, Y.~Zhao, J.~Wan, K.~Zhou, F.~Zhang, Z.~Wang, and K.~Zheng.
\newblock Multi-modal knowledge graphs for recommender systems.
\newblock In {\em CIKM}, pages 1405--1414, 2020.

\bibitem{wsdm21coldstart}
R.~Togashi, M.~Otani, and S.~Satoh.
\newblock Alleviating cold-start problems in recommendation through
  pseudo-labelling over knowledge graph.
\newblock In {\em WSDM}, pages 931--939, 2021.

\bibitem{nips17metaRec}
M.~Vartak, A.~Thiagarajan, C.~Miranda, J.~Bratman, and H.~Larochelle.
\newblock A meta-learning perspective on cold-start recommendations for items.
\newblock In {\em NeurIPS}, pages 6904--6914, 2017.

\bibitem{iclr18gat}
P.~Veličković, G.~Cucurull, A.~Casanova, A.~Romero, P.~Liò, and Y.~Bengio.
\newblock Graph attention networks.
\newblock In {\em ICLR}, 2018.

\bibitem{cikm18ripple}
H.~Wang, F.~Zhang, J.~Wang, M.~Zhao, W.~Li, X.~Xie, and M.~Guo.
\newblock Ripplenet: Propagating user preferences on the knowledge graph for
  recommender systems.
\newblock In {\em CIKM}, pages 417--426, 2018.

\bibitem{www18dkn}
H.~Wang, F.~Zhang, X.~Xie, and M.~Guo.
\newblock Dkn: Deep knowledge-aware network for news recommendation.
\newblock In {\em WWW}, pages 1835--1844, 2018.

\bibitem{kdd19kgnn-ls}
H.~Wang, F.~Zhang, M.~Zhang, J.~Leskovec, M.~Zhao, W.~Li, and Z.~Wang.
\newblock Knowledge-aware graph neural networks with label smoothness
  regularization for recommender systems.
\newblock In {\em KDD}, pages 968--977, 2019.

\bibitem{www19kgcn}
H.~Wang, M.~Zhao, X.~Xie, W.~Li, and M.~Guo.
\newblock Knowledge graph convolutional networks for recommender systems.
\newblock In {\em WWW}, pages 3307--3313, 2019.

\bibitem{tkde17kgesuvey}
Q.~Wang, Z.~Mao, B.~Wang, and L.~Guo.
\newblock Knowledge graph embedding: A survey of approaches and applications.
\newblock {\em TKDE}, 29(12):2724--2743, 2017.

\bibitem{kdd19kgat}
X.~Wang, X.~He, Y.~Cao, M.~Liu, and T.~Chua.
\newblock {KGAT}: Knowledge graph attention network for recommendation.
\newblock In {\em KDD}, pages 950--958, 2019.

\bibitem{sigir19ngcf}
X.~Wang, X.~He, M.~Wang, F.~Feng, and T.~Chua.
\newblock Neural graph collaborative filtering.
\newblock In {\em SIGIR}, pages 165--174, 2019.

\bibitem{aaai19explainable}
X.~Wang, D.~Wang, C.~Xu, X.~He, Y.~Cao, and T.~Chua.
\newblock Explainable reasoning over knowledge graphs for recommendation.
\newblock In {\em AAAI}, pages 5329--5336, 2019.

\bibitem{www20KGPolicy}
X.~Wang, Y.~Xu, X.~He, Y.~Cao, M.~Wang, and T.-S. Chua.
\newblock Reinforced negative sampling over knowledge graph for recommendation.
\newblock In {\em WWW}, pages 99--109, 2020.

\bibitem{sigir20ckan}
Z.~Wang, G.~Lin, H.~Tan, Q.~Chen, and X.~Liu.
\newblock {CKAN}: Collaborative knowledge-aware attentive network for
  recommender systems.
\newblock In {\em SIGIR}, pages 219--228, 2020.

\bibitem{wu2020garg}
S.~Wu, Y.~Zhang, C.~Gao, K.~Bian, and B.~Cui.
\newblock Garg: anonymous recommendation of point-of-interest in mobile
  networks by graph convolution network.
\newblock {\em Data Science and Engineering}, 5(4):433--447, 2020.

\bibitem{nips21scheduler}
H.~Yao, Y.~Wang, Y.~Wei, P.~Zhao, M.~Mahdavi, D.~Lian, and C.~Finn.
\newblock Meta-learning with an adaptive task scheduler.
\newblock {\em NeurIPS}, 2021.

\bibitem{YuQLL20}
H.~Yu, T.~Qian, Y.~Liang, and B.~Liu.
\newblock {AGTR:} adversarial generation of target review for rating
  prediction.
\newblock {\em Data Sci. Eng.}, 5(4):346--359, 2020.

\bibitem{wsdm14per}
X.~Yu, X.~Ren, Y.~Sun, Q.~Gu, B.~Sturt, U.~Khandelwal, B.~Norick, and J.~Han.
\newblock Personalized entity recommendation: A heterogeneous information
  network approach.
\newblock In {\em WSDM}, pages 283--292, 2014.

\bibitem{kdd16cke}
F.~Zhang, N.~J. Yuan, D.~Lian, X.~Xie, and W.~Ma.
\newblock Collaborative knowledge base embedding for recommender systems.
\newblock In {\em KDD}, pages 353--362, 2016.

\bibitem{RecSurvey19}
S.~Zhang, L.~Yao, A.~Sun, and Y.~Tay.
\newblock Deep learning based recommender system: A survey and new
  perspectives.
\newblock {\em ACM Comput. Surv.}, 52(1), 2 2019.

\bibitem{kdd19metapred}
X.~S. Zhang, F.~Tang, H.~H. Dodge, J.~Zhou, and F.~Wang.
\newblock Metapred: Meta-learning for clinical risk prediction with limited
  patient electronic health records.
\newblock In {\em KDD}, pages 2487--2495, 2019.

\bibitem{sigir21UGRec}
X.~Zhao, Z.~Cheng, L.~Zhu, J.~Zheng, and X.~Li.
\newblock {UGRec}: Modeling directed and undirected relations for
  recommendation.
\newblock In {\em SIGIR}, pages 193--202, 2021.

\bibitem{recsys18rkge}
S.~Zhu, Y.~Jie, Z.~Jie, B.~Alessandro, H.~Long{-}Kai, and X.~Chi.
\newblock Recurrent knowledge graph embedding for effective recommendation.
\newblock In {\em RecSys}, pages 297--305, 2018.

\end{thebibliography}
